# Towards a full $w$CDM map-based analysis for weak lensing surveys

D. Zürcher,[1] J. Fluri,[1] V. Ajani [ORCID],[1,*] S. Fischbacher,[1] A. Refregier[1,*] and T. Kacprzak[1,2]

[1] *Department of Physics, Institute for Particle Physics and Astrophysics, ETH Zürich, Wolfgang Pauli Strasse 27, 8093 Zürich, Switzerland*
[2] *Swiss Data Science Center, Paul Scherrer Institute, 5232 Villigen PSI, Switzerland*

**ABSTRACT**

The next generation of weak lensing surveys will measure the matter distribution of the local universe with unprecedented precision, allowing the resolution of non-Gaussian features of the convergence field. This encourages the use of higher-order mass-map statistics for cosmological parameter inference. We extend the forward-modelling based methodology introduced in a previous forecast paper to match these new requirements. We provide multiple forecasts for the $w$CDM parameter constraints that can be expected from stage 3 and 4 weak lensing surveys. We consider different survey setups, summary statistics and mass map filters including wavelets. We take into account the shear bias, photometric redshift uncertainties, and intrinsic alignment. The impact of baryons is investigated and the necessary scale cuts are applied. We compare the angular power spectrum analysis to peak and minima counts as well as Minkowski functionals of the mass maps. We find a preference for Starlet over Gaussian filters. Our results suggest that using a survey setup with 10 instead of 5 tomographic redshift bins is beneficial. Adding cross-tomographic information improves the constraints on cosmology and especially on galaxy intrinsic alignment for all statistics. In terms of constraining power, we find the angular power spectrum and the peak counts to be equally matched for stage 4 surveys, followed by minima counts and the Minkowski functionals. Combining different summary statistics significantly improves the constraints and compensates the stringent scale cuts. We identify the most 'cost-effective' combination to be the angular power spectrum, peak counts and Minkowski functionals following Starlet filtering.

## 1 INTRODUCTION

Weak gravitational lensing (WL) refers to the observed distortions of the shapes of background galaxies that are caused by the gravitational lensing due to the foreground large scale structure (LSS) of the universe. While the WL signal from a single galaxy is on the per cent-level, the statistical signal from many galaxies exhibits a strong dependence on cosmology. Through the shape measurement of millions of galaxies, WL allows the reconstruction of the projected matter distribution of the local universe in an unbiased way. Furthermore, the addition of tomographic information enables one to map the evolution of structure formation (see e.g. Bartelmann & Schneider (2001); Schneider (2006); Kilbinger (2015) for a review). Supported by the scientific results of past and ongoing galaxy surveys such as the Canada France Hawaii Telescope Lensing Survey.[1] (CFHTLenS) (Heymans et al. 2013), the Kilo-Degree Survey[2] (KiDS) (Hildebrandt et al. 2020a; Asgari et al. 2021), the Subaru Hyper Suprime-Cam[3] (HSC) (Hikage et al. 2019), and the Dark Energy Survey[4] (DES) (Troxel et al. 2018; Amon et al. 2022; Secco et al. 2022) WL has established itself as a powerful cosmological probe and is listed as one of the major science goals of future large-scale surveys such as the Legacy Survey of Space and Time[5] (LSST) (Chang et al. 2013), the Nancy Grace Roman Space Telescope[6] (NGRST, formerly WFIRST) (Spergel et al. 2015), and *Euclid*[7] (Laureijs et al. 2011). While the projected matter distribution is well approximated by a homogeneous and isotropic Gaussian random field with zero mean on large scales, the non-linear nature of the gravitational collapse of high-density structures makes the field primarily non-Gaussian on small scales. Past studies relied heavily on two-point statistics such as the real-space two-point correlation function (see e.g. Troxel et al. 2018; Secco et al. 2021) or the angular power spectrum (see e.g. Hikage et al. 2019; Asgari et al. 2021) to extract information from the mass maps. Such two-point statistics are sufficient summary statistics for a Gaussian random field with zero mean, but are unable to extract most of the non-Gaussian information on small scales (Springel, Frenk & White 2006; Yang et al. 2011). With the increased resolution of small scale features in upcoming surveys the extraction of non-Gaussian information from the projected matter distribution becomes even more important. Hence, multiple higher order summary statistics were developed in recent years to target this additional non-Gaussian information. Some important examples include three-point statistics (Takada & Jain 2003; Semboloni et al. 2011; Fu et al. 2014), higher order mass map moments (Van Waerbeke et al. 2013;

Petri, May & Haiman 2016; Patton et al. 2017; Gatti et al. 2020, 2021), Minkowski functionals (Kratochvil et al. 2012; Shirasaki & Yoshida 2014; Petri et al. 2015; Parroni et al. 2020), peak counts (Jain & Van Waerbeke 2000; Dietrich & Hartlap 2010; Kacprzak et al. 2016; Fluri et al. 2018; Martinet et al. 2018; Shan et al. 2018; Ajani et al. 2020; Harnois-Déraps et al. 2020; Jeffrey, Alsing & Lanusse 2021a; Lanzieri et al. 2023), minima counts and voids (Coulton et al. 2020; Davies et al. 2021), Betti numbers (Parroni et al. 2021), the one-point probability distribution function (Barthelemy et al. 2020; Boyle et al. 2021; Martinet et al. 2021), the $\ell_1$-norm (Ajani, Starck & Pettorino 2021), persistent homology (Heydenreich, Brück, Benjamin & Harnois-Déraps, Joachim 2021), wavelet based methods (Allys et al. 2020; Cheng et al. 2020), machine learning based approaches (Gupta et al. 2018; Fluri et al. 2019; Jeffrey et al. 2021a; Fluri et al. 2022) as well as multiscale flows (Dai & Seljak 2023), and field-level inference (Porqueres et al. 2021; Boruah & Rozo 2023; Porqueres et al. 2023). The use of higher order mass map statistics is hindered in practice by the fact that most of them are very difficult to predict from a theoretical framework due to the highly non-linear nature of the gravitational collapse of structures in the late universe. Zürcher et al. (2021) (hereafter Z21) developed an analysis pipeline that circumvents this issue by relying on forward-modelling. The statistics are predicted numerically based on a large suite of *N*-body simulations. The developed framework was shown to recover the input cosmology when using synthetic weak lensing data. Furthermore, the analysis pipeline was successfully applied to the shape catalogue generated from the first three years of data of the DES (Gatti et al. 2020) obtaining state-of-the-art constrains on the structure growth parameter $S_8$ by Zürcher et al. (2022) (hereafter Z22).

The goal of this paper is twofold. First, we improve the methodology developed in Z21 and applied to data in Z22. More specifically we expand the studied parameter space from the matter density $\Omega_{\rm m}$ and amplitude of density fluctuations $\sigma_8$ to a full *w*CDM analysis using the significantly larger *N*-body suite COSMOGRID (Kacprzak et al. 2022), we include a wavelet-based filtering scheme known as the Starlet decomposition (Starck, Fadili & Murtagh 2007) and we transition from a Gaussian Process Regressor to a neural network emulator. Secondly, we investigate the forecasted performance of some mass map summary statistics (specifically the angular power spectrum, peak counts, minima counts and Minkowski functionals) in different survey setups. We study how the statistics compare in terms of constraining power and robustness to galaxy intrinsic alignment (IA). Furthermore, the importance of cross-tomographic information, the number of redshift bins and the choice of the mass map filter scheme is investigated for all the different statistics. Using a set of baryon contaminated *N*-body simulations, we derive how the statistics are affected by baryonic physics and we derive the scale cuts that are necessary to obtain unbiased results in a stage 4 survey setup. Based on our findings we make suggestions for some analysis choices for a stage 4, full *w*CDM analysis using higher order statistics. Lastly, we also explore the potential of different combinations of summary statistics.

This paper starts with a short introduction about weak gravitational lensing, mass mapping, and mass map statistics in Section 2. In Section 3, the used *N*-body simulation suite is presented and the generation of the mock shape catalogue is described. Section 4 introduces how we forward model the mass maps and summary statistics for different cosmologies and the parameter inference. We present the findings of this work in Section 5. Finally, we summarize the main outcomes of this study and make suggestions for future works in Section 6. Further, we show the distribution of the used COSMOGRID simulations in the *w*CDM parameter space in Appendix A and we include a detailed presentation of the inferred parameter constraints in Appendices B and C.

## 2 THEORY

In the following Section, we briefly revisit the essentials about weak gravitational lensing and mass mapping. Furthermore, we introduce the summary statistics studied in this work as well as the Starlet decomposition that we employ to filter the maps.

### 2.1 Weak gravitational lensing

The paths of photons emitted by distant galaxies can be altered by their passage through local deformations of space-time geometry caused by the foreground LSS of the universe. This effect, that is commonly referred to as gravitational lensing, can lead to distortions in the apparent shapes of the galaxies. As such distortions are typically on the per cent level this regime is called weak gravitational lensing (WL). We direct the reader to Bartelmann & Schneider (2001) for a thorough review of the principles of gravitational lensing. WL leads to two distinct changes of the observed shapes of background galaxies; an isotropic magnification $\kappa$ of the galaxy size and an anisotropic stretching $\gamma = \gamma_1 + i\gamma_2$ of the ellipticity of the galaxy. These two quantities can be expressed as a function in spherical coordinates $\kappa(\theta, \varphi)$, $\gamma(\theta, \varphi)$ and are dubbed convergence and cosmic shear, respectively.

Neither the cosmic shear nor the convergence fields are accessible in real-world weak lensing surveys. Instead, only the galaxy ellipticities $e = e_1 + ie_2$ can be measured. They can be regarded as a combination of the intrinsic galaxy shape $e_s$ and the reduced shear $g$

$$e = g + e_s = \frac{\gamma}{1 + \kappa} + e_s \approx \gamma + e_s, \tag{1}$$

where we assumed the weak gravitational limit to approximate $g \approx \gamma$. The intrinsic galaxy shapes $e_s$ are unknown but can be assumed to be random with no preferred orientation, such that $<e_s> = 0$, in the absence of galaxy IA. Therefore, the measured galaxy ellipticities $e$ can be understood as noisy measurements of the cosmic shear

$$\gamma_{\rm obs} = m\gamma + c + e_s, \tag{2}$$

where the noise term $e_s$ is often referred to as shape noise. Since the shape noise contribution is typically larger than the cosmic shear signal by about two orders of magnitude, the statistical potency of the signal must be enhanced by averaging over multiple galaxies in the same region of the sky (Bartelmann & Schneider 2001). The shear bias terms are indicated by $m$ (multiplicative shear bias) and $c$ (additive shear bias). Shear bias is a major systematic in weak lensing measurements, that we discuss in Section 4.2.2. For the remainder of this theoretical introduction we assume $m = 1$, $c = 0$.

### 2.2 Mass mapping

The challenge of calculating the convergence signal $\kappa$ from the noisy, often only partial sky cosmic shear field $\gamma_{\rm obs}$ is referred to as mass mapping and several methods have been developed in the past to address this problem, some of which are explored in Jeffrey et al. (2021b).

We follow a direct, analytical formalism developed in Kaiser, Squires & Broadhurst (1994); Wallis et al. (2017) that is commonly referred to as the spherical Kaiser–Squires (KS) method. Being defined as spherical fields, the convergence and cosmic shear signals can be decomposed in the basis of spherical harmonics $_s Y_{\ell m}(\theta, \varphi)$ with spin-weight $s$ as

$$\gamma(\theta, \varphi) = \sum_{\ell=0}^{\ell_{\max}} \sum_{m=-\ell}^{\ell} {}_2\hat{\gamma}_{\ell m} \, {}_2 Y_{\ell m}(\theta, \varphi), \tag{3}$$

$$\kappa(\theta, \varphi) = \sum_{\ell=0}^{\ell_{\max}} \sum_{m=-\ell}^{\ell} {}_0\hat{\kappa}_{\ell m} \, {}_0 Y_{\ell m}(\theta, \varphi). \tag{4}$$

The series are infinite in $\ell$ and they need to be truncated at an $\ell_{\max}$ in practical applications. We use the `map2alm` routine of the HEALPIX[8] software Gorski et al. (2005) to decompose the fields in spherical harmonic space. We use a pixel resolution of NSIDE = 1024 with a default $\ell_{\max} = 3 \cdot \text{NSIDE} - 1 = 3071$.

It can be shown that both fields can be related to the lensing potential $\phi(\theta, \varphi)$ using the covariant derivative $\eth$ and its adjoint $\bar{\eth}$, through (e.g. Kaiser & Squires (1993a); Hu (2000); Bartelmann (2010)):

$$\gamma = \frac{1}{2}\eth\eth\phi, \tag{5}$$

$$\kappa = \frac{1}{4}(\eth\bar{\eth} + \bar{\eth}\eth)\phi. \tag{6}$$

In spherical harmonics space these relations read

$$_2\hat{\gamma}_{\ell m} = \frac{1}{2}\sqrt{(\ell-1)\ell(\ell+1)(\ell+2)}\,_0\hat{\phi}_{\ell m}, \tag{7}$$

$$_0\hat{\kappa}_{\ell m} = -\frac{1}{2}\ell(\ell+1)\,_0\hat{\phi}_{\ell m}. \tag{8}$$

Hence, we can obtain a relation connecting the convergence and cosmic shear fields in harmonic space

$$_2\hat{\gamma}_{\ell m} = \mathcal{D}_\ell \,_0\hat{\kappa}_{\ell m}, \tag{9}$$

where the kernel $\mathcal{D}_\ell$ is defined as

$$\mathcal{D}_\ell = -\sqrt{\frac{(\ell-1)(\ell+2)}{\ell(\ell+1)}}. \tag{10}$$

### 2.3 Mass map summary statistics

The convergence signal can be decomposed into a curl-free (E-modes) and divergence-free (B-modes) part as $\kappa = \kappa^E + i\kappa^B$ (or in spherical harmonic space $_0\hat{\kappa}_{\ell m} = _0\hat{\kappa}_{\ell m}{}^E + i_0\hat{\kappa}_{\ell m}{}^B$). Gravitational lensing only produces an E-mode signal, while systematic effects like galaxy IAs and also imperfections in the shear calibration pipeline can give rise to B-modes as well. In the presence of a mask, mode mixing occurs, leading to the production of mixed EB-modes and the leakage of E-modes into the B-mode signal and vice-versa. Hence, we only calculate mass map summary statistics from the E-mode part of $\kappa$ to constrain cosmology as it is expected to carry most of the cosmological information. We will drop the E superscript in the following. Further, we note that masking effects are included in the numerical predictions of the summary statistics due to the forward-modelling scheme followed in this study and we do not need to correct for such effects.

In the following, we briefly introduce the summary statistics studied in this work and explain how they are measured from the mass maps.

### 2.3.1 Angular power spectrum

Having the advantage of being accurately predictable from the matter power spectrum using the Limber approximation (Limber 1953), angular power spectra are used as one of the primary ways to extract information from mass maps in WL studies (see e.g. Heymans et al. (2013); Hildebrandt et al. (2017); Troxel et al. (2018); Hikage et al. (2019); Amon et al. (2021); Heymans et al. (2021); Secco et al. (2021)). We calculate the auto- and cross angular power spectra from the tomographic mass maps in the spherical harmonics space as

$$C_{\ell,(i,j)} = \frac{1}{2\ell+1}\sum_{m=-\ell}^{\ell} \hat{\kappa}_{\ell m, i} \cdot {}_0\hat{\kappa}^*_{\ell m, j}, \tag{11}$$

using the HEALPIX routine `anafast` (Zonca et al. 2019). The indices $i$ and $j$ indicate the tomographic bins. We measure the angular power spectra of the maps using a range of $\ell \in [1, 2048]$ and 25 square-root-spaced bins.

### 2.3.2 Higher order statistics

Given the decomposition in spherical harmonic space of two tomographic mass maps $_0\hat{\kappa}_{\ell m, i}$ and $_0\hat{\kappa}_{\ell m, j}$ from two tomographic bins $i$ and $j$ we produce maps as

$$\kappa_{i,j}(\theta, \varphi) = \sum_{\ell=0}^{\ell_{\max}} \sum_{m=-\ell}^{\ell} \sqrt{_0\hat{\kappa}_{\ell m, i}} \sqrt{_0\hat{\kappa}_{\ell m, j}} \,_0 Y_{\ell m}(\theta, \varphi), \tag{12}$$

using the HEALPIX routine `alm2map`. Taking $i = j$ results in the reconstruction of the tomographic mass map corresponding to tomographic bin $i$, while a cross-tomographic map is obtained when choosing $i \neq j$. We utilize a multiscale approach in this study in order to extract non-Gaussian information from features at different scales. Such an approach was shown to extract more information from the convergence maps compared to a single-scale approach (see e.g. Fluri et al. (2018); Ajani et al. (2020); Z21). Hence, we subsequently filter the maps with a set of filters before the calculation of the higher-order statistics. We consider two types of filters: (1) a scale-space filtering using a set of Gaussian filters and (2) a wavelet decomposition using starlet wavelets. In the following Section, we introduce how these filtered map versions are created from the original map. We follow the methodology and notation from Starck et al. (2006a).

Conceptually, we produce $k$ filtered versions $c_s(\theta, \varphi)$ of the original map $\kappa(\theta, \varphi)$ by performing convolutions with scaling functions $\phi_s(\theta, \varphi)$, where $s$ runs over the different filter scales

$$c_0 = \phi_0 * \kappa,$$
$$c_1 = \phi_1 * \kappa,$$
$$\dots$$
$$c_k = \phi_k * \kappa.$$

In practice, the convolution is executed in spherical harmonic space as

$$\hat{c}_s(\ell, m) = \sqrt{\frac{4\pi}{2\ell+1}}\hat{\phi}_s(\ell, 0)\hat{\kappa}(\ell, m). \tag{13}$$

Note that all spherical harmonic coefficients $\hat{\phi}_s(\ell, m \neq 0)$ vanish as we only consider axis-symmetric scaling functions $\phi_s$ that do not depend on $\varphi$. The definition of the low-pass filters

$$\hat{H}_s(\ell, m) = \sqrt{\frac{4\pi}{2\ell+1}}\hat{h}_s(\ell, m) \tag{14}$$

$$= \begin{cases} \frac{\hat{\phi}_s(\ell,m)}{\hat{\phi}_{s-1}(\ell,m)} & \text{if } \ell < \ell_s \text{ and } m = 0 \\ 0 & \text{else} \end{cases} \tag{15}$$

allows to calculate the smoothed maps $\hat{c}_s(\ell, m)$ in a more efficient, recursive fashion. $\ell_s$ indicates the characteristic cut-off frequency corresponding to scale $s$ such that

$$\phi_s(\theta, \varphi) = \sum_{\ell=0}^{\ell_s} \hat{\phi}_s(\ell, 0) \,_0 Y_{\ell 0}(\theta, \varphi). \tag{16}$$

The maps $\hat{c}_s$ can then be calculated recursively as $\hat{c}_s = \hat{H}_s(\ell, m)\hat{c}_{s-1}$.

In the case of the scale-space filtering scheme, we use scaling functions $\phi_s(\theta, \varphi)$ that are Gaussian in real space and characterized by their full-width-at-half-maximum (FWHM). The corresponding spherical harmonic coefficients are

$$\hat{\phi}_s(\ell, 0) = \exp\left(-\frac{1}{2}\ell(\ell+1)\sigma^2\right), \tag{17}$$

where

$$\sigma = \frac{\text{FWHM}}{60} \frac{\pi}{180} \frac{1}{2\sqrt{2\ln(2)}}. \tag{18}$$

In our analysis, we use 12 such scaling functions with FWHM $\in$ [31.6, 29.0, 26.4, 23.7, 21.1, 18.5, 15.8, 13.2, 10.5, 7.9, 5.3, 3.3] arcmin. We adapt the same FWHMs as used previously in Fluri et al. (2018); Z21; Zürcher et al. (2022) but we change the smallest scale from 2.7 to 3.3 arcmin. We will refer to this filtering scheme as the GAUSS scheme. The resulting spherical harmonic coefficient $\hat{\phi}_s(\ell, 0)$ are shown in the left-most panel of Fig. 1. Note that there is no cut-off scale for the Gaussian filters, resulting in $\ell_s = \infty$.

In the second approach, we extract information from the wavelet decomposition $w_s(\theta, \varphi)$ of $\kappa(\theta, \varphi)$, where the index $s$ runs over a set of different filter scales once again. Having decomposed the original map into smoothed versions $c_s$ as described earlier the wavelet decomposition can be obtained as

$$w_s(\theta, \varphi) = c_{s-1}(\theta, \varphi) - c_s(\theta, \varphi). \tag{19}$$

The wavelet decomposition $w_s(\theta, \varphi)$ of $\kappa(\theta, \varphi)$ can be understood as a version of $\kappa$ that was smoothed with a band-pass filter $\hat{\psi}_s(\ell, m) = \hat{\phi}_{s-1}(\ell, m) - \hat{\phi}_s(\ell, m)$. The filters $\hat{\psi}_s(\ell, m)$ are commonly referred to as wavelets. Wavelets have emerged as popular and powerful tools in cosmology in the past decades (Starck, Pantin & Murtagh 2002) for example for extracting non-Gaussianities from the Cosmological Microwave Background (Aghanim & Forni 1999; Barreiro & Hobson 2001; Starck, Aghanim & Forni 2004; Vielva et al. 2004), reconstructing the primordial power spectrum (Fang & Feng 2000) or reconstruction of WL mass maps (Starck, Pires & Réfrégier 2006b; Lanusse 2015; Lanusse, F. et al. 2016; Peel, Lanusse & Starck 2017b), among other applications. While in principle many functions would make for reasonable scaling function to construct wavelets, Box splines of order 3 have emerged as a very popular choice in astronomy and cosmology (Starck, Murtagh & Fadili 2015). Thus, the scaling functions read

$$\hat{\phi}_s(\ell, 0) = \frac{3}{2} B_3\left(\frac{2\ell}{\ell_s}\right), \tag{20}$$

$$B_3(x) = \frac{1}{12}(|x-2|^3 - 4|x-1|^3 + 6|x|^3 - 4|x+1|^3 + |x+2|^3). \tag{21}$$

The scaling functions are characterized by their cut-off scales $\ell_s$. The larger $\ell_s$, the more small scale information is included. The isotropic wavelet transform obtained using the above isotropic $B_3$-spline scaling functions is also known under the name of starlet transform (Starck et al. 2007). Because of this scaling function, the resulting wavelet is a compensated function with compact support in [−2, 2] × [−2, 2] (Starck et al. 2015). It can be fully defined by its symmetric filters from the filter bank and its dictionary is very well adapted to astronomical applications. It has also been shown to be a powerful tool in extracting cosmological information in an ideal, planar setting (Ajani et al. 2021) and to present advantageous characteristics such as a more diagonal covariance matrix in the context of peak counts, compared to Gaussian filters (Ajani et al. 2020).

We consider two different sets of cut-off frequencies for the starlets in this study. First, we use the originally proposed dyadic scheme in which each cut-off frequency $\ell_s$ can be obtained as $\ell_s = 2\ell_{s-1}$. The largest cut-off frequency is set to $4 \cdot$ NSIDE. This results in $\ell_s \in$ [8, 16, 32, 64, 128, 256, 512, 1024, 2048, 4096]. We will refer to this filtering scheme as the DYADIC scheme. The wavelet formalism is not restricted to a dyadic spacing of the filters and alternative wavelet transforms that allow filtering at any intermediate scales are possible. Therefore, for the purposes of this study, we also construct a second set of cut-off frequencies $\ell_s \in$ [1000, 1221, 1491, 1821, 2223, 2715, 3315, 4048, 4943, 6035, 7370, 8999, 10988, 13418, 16384] with a larger number of cut-off frequencies and more of them concentrated at smaller scales (the frequencies are still spaced linearly in log space but such that $\ell_s = 1.22\,\ell_{s-1}$). We expect that this choice of cut-off frequencies leads to an increase in the amount of extracted cosmological information as small-scale structures, that carry most of the non-Gaussian information, are better resolved. We will refer to this filtering scheme as the LOG scheme. We have chosen the largest cut-off frequency of 16 384 such that the stacked profiles around local maxima identified from the mass maps are the most similar to the ones found for maps smoothed with a Gaussian kernel with FWHM = 3.3 arcmin. This was done to assure that the smallest features accessible in both approaches are approximately of the same scale. The resulting Fourier responses of the wavelets $\hat{\psi}_s(\ell, m)$ corresponding to these two schemes are shown in the middle and right-most plot in Fig. 1.

We make the developed software used to calculate the starlet decompositions of the spherical maps in this project publicly available at eSD[9]. The core functions of eSD are written in C/C++ and taken from CosmoStat[10] (Starck et al. 2021) and Sparse2D[11] (Starck, Murtagh & Fadili 2010). eSD includes an easy-to-use PYTHON wrapper around the just mentioned C++ core routines and is installable out-of-the-box.

In the following Section, we introduce the higher-order statistics investigated in this study. Namely local extrema (peaks and minima) and Minkowski functionals. We calculate all statistics in all three filtering schemes.

### 2.3.3 Local extrema

The number counts of local maxima (peaks) on pixelized maps as a function of either their convergence or signal-to-noise (SNR, defined as SNR $= \kappa / <\sigma_\kappa>$ here) was demonstrated to be an excellent higher-order statistic that is complementary to two-point statistics (see e.g. Reblinsky et al. (1999); Yang et al. (2011); Liu & Haiman (2016); Z21) and was previously used to constrain cosmology from observed data (see e.g. Kacprzak et al. (2016); Harnois-Déraps et al. (2021); Zürcher et al. (2022)). Initially introduced to detect massive structures from the weak lensing signal, lensing peaks trace over-dense regions of the LSS in an unbiased way (Tyson, Valdes & Wenk 1990; Miralda-Escude 1991; Kaiser & Squires 1993b; Schneider 1996).

---

[9] https://cosmo-gitlab.phys.ethz.ch/cosmo_public/eSD
[10] https://github.com/CosmoStat/cosmostat
[11] https://github.com/cosmostat/sparse2d

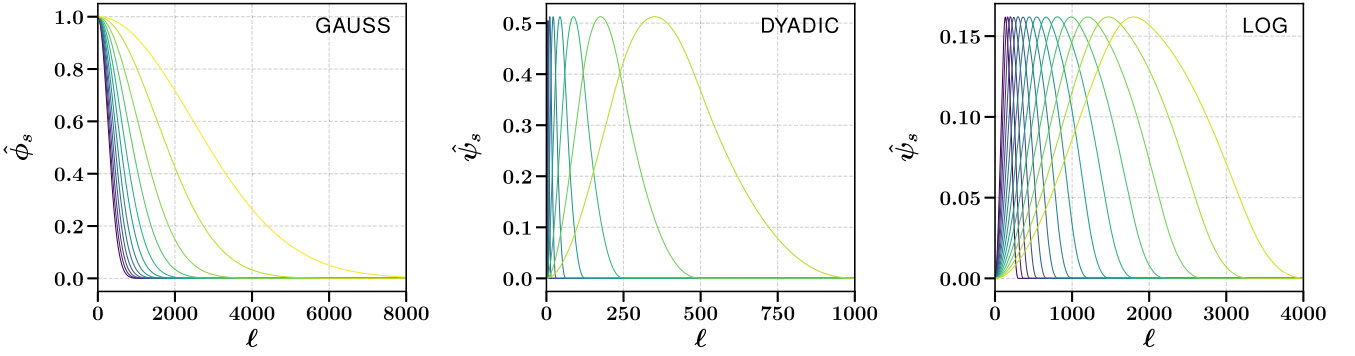

**Figure 1. Left:** the scaling functions used in the GAUSS scheme. These are used to smooth the mass maps before the extraction of information and correspond to Gaussian kernels with FWHM ∈ [31.6, 29.0, 26.4, 23.7, 21.1, 18.5, 15.8, 13.2, 10.5, 7.9, 5.3, 3.3] arcmin in real space. **Middle:** the starlet wavelets used to filter the mass maps in the DYADIC scheme. The cut-off scales of the filters are $\ell_s \in$ [8, 16, 32, 64, 128, 256, 512, 1024, 2048, 4096]. **Right:** the corresponding starlet wavelets used in the LOG scheme with $\ell_s \in$ [1000, 1221, 1491, 1821, 2223, 2715, 3315, 4048, 4943, 6035, 7370, 8999, 10988, 13418, 16384].

Targeting the complementary, underdense regions of mass maps, the number counts of local minima offer a promising alternative to peak counts. While the extracted information was demonstrated to be correlated with the information from peak counts, there are some indications that local minima are less susceptible to systematic biases from baryonic physics (Coulton et al. 2020; Z21). Additionally, the study of underdense regions might offer insights into physics beyond the standard model such as modified gravity (Baker et al. 2018; Paillas et al. 2019).

We regard a pixel of the filtered mass maps as a peak/minimum if its convergence value is higher/lower than all its neighbouring pixels. Note however, that the definition of a lensing peak/minimum is not unique. Other studies use the local maxima on aperture-mass maps to identify lensing peaks for example Kacprzak et al. (2016); Martinet et al. (2018); Harnois-Déraps et al. (2021). All recorded peaks/minima are then binned as a function of their convergence value into 15 linearly spaced bins. To suppress shot-noise contributions and make the likelihood more Gaussian we adjust the edges of the first and last bins such that at least 30 peaks/minima are found in both of these bins in all simulations. Additionally, we restrict ourselves to peaks with SNR ≤ 4.0 (and minima with SNR ≥ −4.0) as such strong structures were found to be potentially biased by source clustering (Kacprzak et al. 2016; Harnois-Déraps et al. 2021; Zürcher et al. 2022).

### 2.3.4 Minkowski functionals

By defining the excursion sets $Q_{\mathrm{SNR}} = \{f : f \geq \mathrm{SNR} \cdot \sigma_f\}$ the Minkowski functionals of a filtered mass map $f$ (being either $c_s$ or $w_s$ depending on the used filter scheme) can be calculated as a function of SNR as

$$V_0(\mathrm{SNR}) = \frac{1}{Q_{\mathrm{SNR}}} \int_{Q_{\mathrm{SNR}}} \Theta(f(\vec{k}) - \mathrm{SNR} \cdot \sigma_f) \, \mathrm{d}x \, \mathrm{d}y, \tag{22}$$

$$V_1(\mathrm{SNR}) = \frac{1}{4Q_{\mathrm{SNR}}} \int_{Q_{\mathrm{SNR}}} \delta(f(\vec{k}) - \mathrm{SNR} \cdot \sigma_f) \sqrt{(\partial_x f)^2 + (\partial_y f)^2} \, \mathrm{d}x \, \mathrm{d}y,$$

$$V_2(\mathrm{SNR}) = \frac{1}{2\pi \, Q_{\mathrm{SNR}}} \int_{Q_{\mathrm{SNR}}} \delta(f(\vec{x}) - \mathrm{SNR} \cdot \sigma_f)$$
$$\times \frac{2\partial_x f \partial_y f \partial_x \partial_y f - (\partial_x f)^2 \partial_y^2 f - (\partial_y f)^2 \partial_x^2 f}{(\partial_x f)^2 + (\partial_y f)^2} \, \mathrm{d}x \, \mathrm{d}y,$$

where $\sigma_f$ denotes the standard deviation of the filtered mass map and $\delta$ and $\Theta$ the Dirac delta and Heaviside-step functions, respectively (Petri et al. 2013). We calculate the derivatives numerically on the pixel level. The Minkowski functionals capture the global topology of a continuous, stochastic field and were shown to probe deviations from Gaussianity (Mecke, Buchert & Wagner 1993). The three functionals $V_0$, $V_1$, and $V_2$ can be interpreted as describing the area, the perimeter and the Euler characteristic of the excursion sets (Vicinanza et al. 2019). We measure each Minkowski functional from the mass maps as a function of SNR in the range [−4.0, 4.0] divided in 10 linearly spaced bins. We are not aware of a study that investigates on the impact of source clustering on the Minkowski functionals. Hence, we adapt the same SNR range as for the other statistics.

## 3 DATA

In this section, we explain the production and analysis choices made for the two main ingredients required in this study: (1) a mock weak lensing shape catalogue with realistic galaxy shapes and redshifts, (2) a large suite of dark-matter-only $N$-body simulations covering the studied parameter space.

### 3.1 Mock survey

We study and compare the performance of different mass map summary statistics in three different scenarios that we dub STAGE 3, STAGE 4 (5 Z-BINS), and STAGE 4 (10 Z-BINS). A summary of the key survey properties of the different setups is presented in Table 1.

We consider a DES-like stage 3 weak lensing shape catalogue in the STAGE 3 setup. We assume a survey area of 5'000 $\mathrm{deg}^2$ and a galaxy density $\mathrm{n_{gals}}$ of 10 galaxies $\mathrm{arcmin}^{-2}$, which corresponds to the values expected for the full DES survey (Dark Energy Survey Collaboration 2005). The positions of the source galaxies on the sky are drawn randomly within the survey area until the target galaxy density is reached. The intrinsic ellipticities $e_s$ of the galaxies are drawn from the probability distribution

$$\mathrm{Prob}(e_s) \propto (e_s + 0.01)^{-4}[1 - \exp(-23e_s^4)], \tag{23}$$

**Table 1.** Summary of the key properties of the three different mock survey setups Stage 3, Stage 4 (5 z-bins), and Stage 4 (10 z-bins). The last row indicates the adapted parameter values for the global Smail distributions from which the galaxy redshifts are drawn.

|  | Stage 3 | Stage 4 (5 z-bins) | Stage 4 (10 z-bins) |
|---|---|---|---|
| Area [deg$^2$] | 5'000 | 14'300 | 14'300 |
| $n_{gals}$ [arcmin$^{-2}$] | 10 | 28 | 28 |
| # tomo. bins | 5 | 5 | 10 |
| $n(z)$ | $\alpha = 1.5$ | $\alpha = 2.0$ | $\alpha = 2.0$ |
|  | $\beta = 1.1$ | $\beta = 0.68$ | $\beta = 0.68$ |
|  | $z_0 = 0.44$ | $z_0 = 0.11$ | $z_0 = 0.11$ |

and the ellipticity components are obtained as

$$e_{s,1} = \Re[e_s \exp(i\phi)],$$
$$e_{s,2} = \Im[e_s \exp(i\phi)], \tag{24}$$

where the angle $\phi$ is drawn uniformly from the interval $[0, 2\pi[$. This corresponds to the same choice made in Z21. The functional form of the distribution in equation (23) was proposed by Bruderer et al. (2016) and fit to the observed distribution of galaxy ellipticities measured by Troxel et al. (2018). We note that such a mock shape catalogue reproduces the contribution of the shape noise to the total galaxy shape distribution but does not include the contribution from the weak lensing signal. The weak lensing signal is added separately. As it will be explained in Section 4.1, the cosmological signal is obtained first at the level of the convergence map defined in equation (26), then it is converted to shear so that also the shape noise component can be added. Finally, the total simulated shear (signal plus noise) is converted to a noisy simulated convergence map.

The redshifts of the individual galaxies are drawn from a Smail distribution (Smail et al. 1995)

$$n(z) \propto z^\alpha \exp\left(-\left[\frac{z}{z_0}\right]^\beta\right), \tag{25}$$

that is characterized by the three parameters $\alpha$, $\beta$, and $z_0$. The values chosen for $\alpha$, $\beta$, and $z_0$ for the Stage 3 setup correspond to the values adopted in Z21 that were fit to the redshift distribution observed by Troxel et al. (2018), but with $z_0$ altered to account for the increased survey depth of the completed DES. The galaxies are then subdivided into 5 tomographic bins according to their redshifts, following the scheme introduced in Amara & Réfrégier (2007). The used strategy assures that each tomographic bin contains the same number of galaxies. The resulting tomographic redshift distributions are presented in the left-most plot in Fig. 2.

We use the same strategies to produce the mock surveys for the Stage 4 (5) and Stage 4 (10) setups, but we change the survey properties to reflect what is expected from a completed stage 4 weak lensing survey. We consider an LSST-like setup and model our mock survey according to the recommendations defined by The LSST Dark Energy Science Collaboration (2018). Hence, we increase the survey area and the galaxy density $n_{gals}$ to 14'300 deg$^2$ and 28 galaxies arcmin$^{-2}$ (accounting for blending), respectively. The ellipticities of the galaxies are drawn from the same probability distribution as in the Stage 3 setup. While the global redshift distribution of the galaxies still follows a Smail distribution it is considerably deeper than in the Stage 3 setup (see Table 1). We consider two different scenarios for the stage 4 setups: Stage 4 (5 z-bins) and Stage 4 (10 z-bins) with the source galaxies being subdivided into 5 and 10 tomographic bins, respectively. The corresponding tomographic redshift distributions are shown in the middle and right-most panels in Fig. 2.

## 3.2 CosmoGrid

In order to accurately predict the mass map statistics at different cosmologies we require a large suite of dark-matter-only $N$-body simulations that sample the desired parameter space. For this purpose, we use part of the CosmoGrid simulation suite introduced in Kacprzak et al. (2022) and previously used successfully to infer cosmology from the KiDS-1000 survey using a Graph Convolutional Neural Network (GCNN) (Fluri et al. 2022).

Initially intended for machine-learning applications the CosmoGrid simulations sample the $w$CDM parameter space spanned by the total matter density $\Omega_m$, baryon density $\Omega_b$, amplitude of density fluctuations $\sigma_8$, scalar spectral index $n_s$, Hubble constant $H_0$, and the equation-of-state parameter of the dark energy component $w$. The 6D $w$CDM space is sampled in 2'500 locations with seven fully-independent simulations each. Additionally, each simulation contains three massive neutrino species that assume a degenerate mass-hierarchy with a mass of $m_v = 0.02$ eV per species. The light neutrino species are modelled as a relativistic fluid in the simulations (Tram et al. 2019). The dark energy density $\Omega_\Lambda$ is adjusted in each simulation to achieve a flat geometry. The simulations are run using the full-tree, GPU-accelerated $N$-body code PkdGrav3 (Potter, Stadel & Teyssier 2017).

The original CosmoGrid simulation suite contains more simulations than needed to reach the precision required for this project and uses very broad priors on the parameters. The sampled parameter space can be divided into a densely sampled inner parameter region (tighter priors) and an outer, more sparsely sampled region. In this study, we use $\sim 25$ per cent of the simulations sampling the inner region, which requires us to put tighter priors on the parameters in the inference process but allows us to have a dense sampling of the studied region without having to run a computationally unfeasible number of mass map simulations. The distribution of the resulting 315 locations in the $\Omega_m - \sigma_8$ plane is shown in Fig. 3. As we additionally study the IA parameters $A_{IA}$ and $\eta$, we need to extend the grid by two additional dimensions. Since the locations of the CosmoGrid simulations are distributed according to a Sobol sequence we can readily extend the dimensionality of the sampled space without losing the sampling properties of the original distribution. The modelling of the IA signal in the simulations is described in Section 4.2.3. The sampled ranges of all parameters, that also dictate the priors used in the inference process, are listed in Table 2. Note that there is an additional prior in the $\Omega_m - \sigma_8$ plane that is motivated by the degeneracy of the two parameters in weak lensing studies and that is not included in Table 2 but shown in Fig. 3 (black box). The distribution of the simulation locations in the full parameter space is presented in Appendix A.

Additionally, we require a large amount of simulations at a fixed cosmology to build the covariance matrices for the different statistics. We use the 200 fiducial simulations of the CosmoGrid suite for this. The location of the fiducial simulations in the $\Omega_m - \sigma_8$ plane is indicated by the black star in Fig. 3. The fiducial setting corresponds to $\Omega_m = 0.26$, $\sigma_8 = 0.84$, $\Omega_b = 0.0493$, $n_s = 0.9649$, $H_0 = 67.36$, $w = -1.0$, $A_{IA} = 0.0$, and $\eta = 0.0$.

PkdGrav3 requires a lookup table of accurate transfer functions that is also used to generate the initial conditions at $z = 99$. This lookup table is calculated using class (Lesgourgues 2011) and transformed into the $N$-body gauge using CONCEPT (Dakin et al.

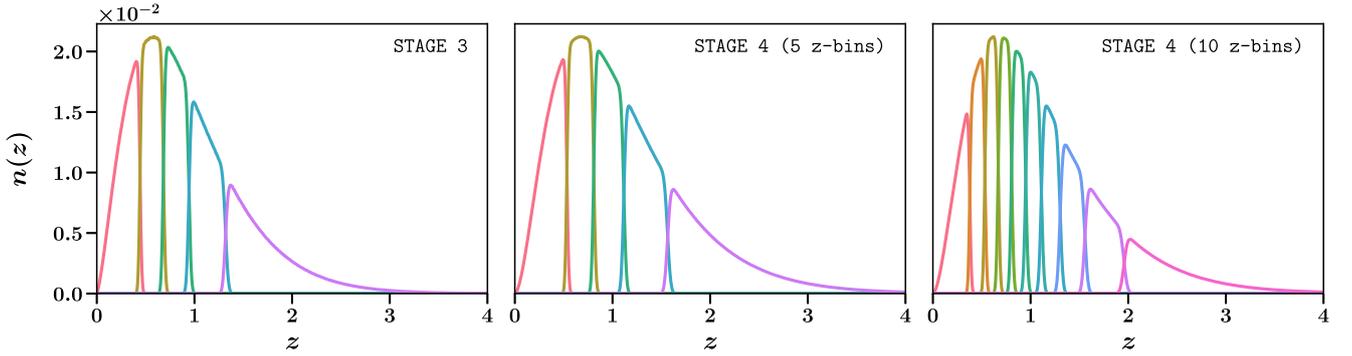

**Figure 2.** Normalized, tomographic redshift distributions of the source galaxies in the produced mock weak lensing surveys. The three plots show the distributions for the three different survey setups STAGE 3, STAGE 4 (5 Z-BINS), and STAGE 4 (10 Z-BINS) from left to right. The global redshift distributions correspond to Smail distributions.

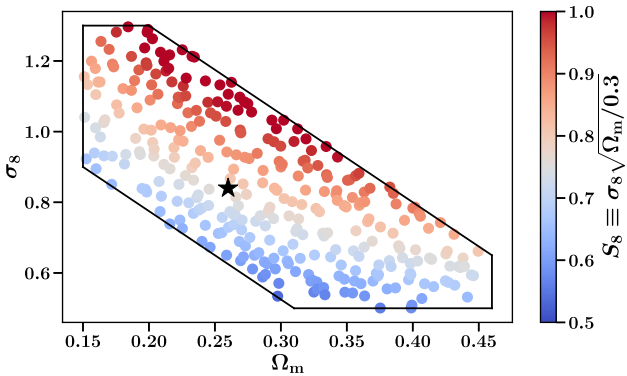

**Figure 3.** Distribution of the 315 locations of the used COSMOGRID simulations in the $\Omega_{\rm m} - \sigma_8$ plane. The colour indicates the $S_8$ value at the parameter location. The location of the fiducial simulations is indicated by the star and the black border indicates the prior in the $\Omega_{\rm m} - \sigma_8$ plane used in the inference step.

**Table 2.** Priors used in the inference step. $\mathcal{U}$ denotes a flat prior with the indicated lower and upper bounds, while $\mathcal{N}$ denotes a normal prior with the indicated mean and scale. Note that there is an additional prior in the $\Omega_{\rm m} - \sigma_8$ plane that is shown in Fig. 3. The priors on $m$ and $\Delta_z$ follow Krause & Eifler (2017).

| Parameter | Prior |
| --- | --- |
| $\Omega_{\rm m}$ | $\mathcal{U}(0.15, 0.45)$ |
| $\sigma_8$ | $\mathcal{U}(0.5, 1.3)$ |
| $\Omega_{\rm b} \times 10^2$ | $\mathcal{U}(4.0, 5.0)$ |
| $n_{\rm s}$ | $\mathcal{U}(0.93, 1.0)$ |
| $H_0$ | $\mathcal{U}(65.0, 75.0)$ |
| $w$ | $\mathcal{U}(-1.25, -0.75)$ |
| $A_{\rm IA}$ | $\mathcal{U}(-2.0, 2.0)$ |
| $\eta$ | $\mathcal{U}(-5.0, 5.0)$ |
| $m_i \times 10^2$ | $\mathcal{N}(0.0, 0.5)$ |
| $\Delta_{z, i} \times 10^2$ | $\mathcal{N}(0.0, 0.5)$ |

2019). The particle positions are then evolved forward to $z = 4$ in 70 time steps. Subsequently, another 70 time steps are taken from $z = 4$ to $z = 0$. All time steps are equally distributed in proper time. The remaining precision parameters of PKDGRAV3 were set to their fiducial values. In all simulations a unit box with a side length of $900 \, {\rm Mpc} \, {\rm h}^{-1}$ filled with $832^3$ particles is used. In order to cover the required redshift range up to $z = 3.5$ the unit box is replicated multiple times. We note, that the redshift distribution of the stage 4 survey setups has a small, but non-negligible tail that reaches beyond $z = 3.5$. Hence, we miss some of the galaxies that are expected to be observed in a stage 4 survey. Therefore, the constraining power of a real stage 4 survey might be slightly higher than predicted in this forecast. However, this small discrepancy does not affect the validity of our results. Using the lightcone mode of PKDGRAV3, spherical particle density maps with a pixel resolution of $\text{NSIDE} = 2048$ are produced. The HEALPIX software is used to pixelize the unit sphere. The resolution of the maps is downgraded to $\text{NSIDE} = 1024$ in the mass mapping step due to memory constraints (see Section 2.2). The baryonic feedback model used to incorporate effects arising from baryonic physics into the simulations additionally requires a halo catalogue. At each time step the built in Friend-of-Friend (FoF) halo finder of PKDGRAV3 is used to produce such a catalogue identifying halos using a linking length of one fifth of the mean particle separation. We refer the reader to Section 5.6 for details about the baryonification of the simulations.

## 4 METHOD

We compare the performance of different higher-order mass map statistics. We do so following a forward modelling approach. For this purpose, we produce realistic spherical mass maps for different cosmologies based on a mock shape catalogue and a suite of dark-matter-only $N$-body simulations. Measuring the different summary statistics from these simulated mass maps allows us to make predictions of the statistics at different cosmologies. In the following Section, we describe how we simulate realistic mass maps from the $N$-body simulations and incorporate systematic effects. Furthermore, we explain how the emulator was set up and how we infer the parameter constraints.

Our methodology builds on Z21; Zürcher et al. (2022). In particular we use the same software and release an updated, more user-friendly version of estats[12] that is available for download from the PyPi[13] as part of this work. The complete codebase used to produce the results presented in this work is publicly available (see NGSF[14]) to assure the reproducibility of these results.

### 4.1 Mass map forward modelling

The UFalcon[15] software is used to combine the particle shells obtained from the different PKDGrav3 simulations into spherical, full-sky mass maps. The UFalcon software developed and described in detail in Sgier et al. (2019) makes use of the Born approximation to avoid full ray-tracing. The Born approximation was shown to be sufficiently accurate for stage 3 as well as stage 4 angular power spectrum analyses (Petri, Haiman & May 2017a).

The spherical density contrast shells $\delta$ of finite redshift thickness $\Delta_{z_b}$ that are obtained from the PKDGrav3 particle shells are converted into spherical mass maps $\kappa$ according to

$$\kappa(\hat{n}) \approx \frac{3\Omega_m}{2} \sum_b W_b^{n(z)} \int_{\Delta_{z_b}} \frac{dz}{E(z)} \delta\left(\frac{c}{H_0} \mathcal{D}(z)\hat{n}, z\right), \quad (26)$$

where $\hat{n}$ indicates the direction on the unit sphere, $H_0$ the Hubble constant, c the speed of light, $\mathcal{D}(z_1, z_2)$ is the dimensionless, comoving distance between redshift $z_1$ and $z_2$, and the function $E(z)$ is defined as

$$d\mathcal{D} = \frac{dz}{E(z)}. \quad (27)$$

The shells are weighted according to the distribution $n(z)$ of the source galaxies in the mock survey using the weights

$$W_b^{n(z)} = \frac{\int_{\Delta_{z_b}} dz \int_z^{z_t} dz' \frac{n(z')}{a(z)E(z)} \frac{\mathcal{D}(0,z)\mathcal{D}(z,z')}{\mathcal{D}(0,z')}}{\left(\int_0^{z_t} dz\, n(z)\right)\left(\int_{\Delta_{z_b}} \frac{dz}{E(z)}\right)}. \quad (28)$$

Subsequently, we cut out an adequate area from the full-sky mass maps using the corresponding mock survey mask. Since none of the studied survey setups covers the whole sky we rotate the full-sky mass maps to obtain multiple mass map realizations from a single PKDGrav3 simulation (8 for the stage 3 setup and 2 for the two stage 4 setups). Lastly, as the convergence map of equation (26) only contains the cosmological signal, realistic shape noise needs to be included in the simulated mass maps. The addition of the noise is performed at the shear level. So first, the simulated mass maps $\kappa$ are converted to a shear signal $\bar{\gamma}$ using spherical KS (see Section 2.2), then the noise obtained by drawing from the mock survey according to equation (23) is added as follows

$$\bar{\gamma}_{sim} = \vec{e}_{noise} + \bar{\gamma} = \frac{\sum_{j=1}^{N_{pix}} \vec{e}_{s,j} \exp(i\phi_j)}{N_{pix}} + \bar{\gamma}, \quad (29)$$

where the sum runs over all galaxies in the mock survey that are located in the corresponding pixel $N_{pix}$ and the angles $\phi_j$ are drawn uniformly from the interval $[0, 2\pi[$. Finally, spherical KS is used once again to transform the fully forward modelled shear signal $\bar{\gamma}_{sim}$ to a mass map $\kappa_{sim}$. The simulated mass map $\kappa_{sim}$ now includes a realistic, statistically equivalent shape noise signal as the mock survey. Additionally, mode mixing effects caused by the survey mask are included in the simulated mass maps due to the use of spherical KS in the generation procedure and are hence forward propagated into the predictions of the summary statistics.

### 4.2 Modelling of systematic effects

Weak lensing studies are susceptible to a variety of systematic effects. Leaving these effects unaccounted for can result in significant biases in the values of inferred cosmological parameters in weak lensing studies. A list of the most prominent systematic effects includes baryonic physics, photometric redshift uncertainty, multiplicative shear bias, as well as galaxy IA. We build on the methodology of Z21; Zürcher et al. (2022) to include photometric redshift uncertainty, multiplicative shear bias and galaxy IA in our analysis. We do not include baryonic feedback effects in the analysis but we investigate the influence of such effects on the different summary statistics using a set of 'baryonified' simulations. We also derive the necessary scale cuts that would be needed to alleviate the impact of such effects on the cosmological parameter constraints (see Section 5.6).

Source clustering is another systematic effect that is known to potentially bias the outcomes of weak lensing studies that use information from local extrema (see e.g. Kacprzak et al. 2016; Harnois-Déraps et al. 2021; Zürcher et al. 2022). Since the strength of the lensing signal depends on the arrangement of the lens and the source galaxies along the line of sight, the lensing signal in a certain direction on the sky is sensitive to the local distribution of the sources in redshift. The presence of a massive foreground structure like a galaxy cluster alters the local redshift distribution of galaxies in its direction as compared to the average distribution in the field and hence weakens the strength of the lensing signal in that direction of the sky. These local changes in the redshift distribution that are present in real world data, are not accounted for in our simulations as the positions of the source galaxies are kept fixed on the sky while the distribution of the dark matter varies depending on the seed used in the PKDGrav3 simulation. This discrepancy can lead to systematic biases when inferring cosmological constraints from observed data and either needs to be forward modelled in the simulations or alleviated by applying appropriate scale cuts. Since we do not have observed data available in this simulation-based study we do not investigate on the strength of this effect, but we apply the same scale cuts as used in Z22 to remove the parts of the data vectors that are potentially affected by this effect, that is, we do not include peaks with a SNR > 4.0 nor minima with SNR < −4.0.

Furthermore, the presence of massive neutrinos has been shown to lead to a systematic change in the detected number of local maxima from mass maps (Fong et al. 2019; Li et al. 2019; Ajani et al. 2020). Since the sum of the massive neutrino masses is kept constant in the PKDGrav3 simulations this could potentially lead to biases when applied to observed data. However, since the change is only significant for high SNR peaks (SNR $\gtrsim$ 4.0) the scale cuts already applied to account for source clustering are enough to alleviate the impact of massive neutrinos (Fong et al. 2019).

In the following Section, we describe how photometric redshift uncertainty, multiplicative shear bias and galaxy IA are incorporated in the analysis.

#### 4.2.1 Photometric redshift uncertainty

To realistically forward model a galaxy survey the tomographic distributions of the source galaxies are required. As it is currently not feasible to measure the redshifts of the individual galaxies spectroscopically in large-scale WL surveys the redshift distributions are inferred photometrically. Although a lot of progress was made in the past years, the photometric determination of the redshift distributions remains challenging. Since inaccuracies in these distributions can lead to biases in the inferred cosmological constraints it is vital to take the uncertainty of the redshift distributions into account (see e.g. Huterer et al. 2006; Choi et al. 2016; Hildebrandt et al. 2020b).

We describe the uncertainty of the redshift distributions $n_i(z)$ by a linear shift described by the parameters $\Delta_{z,i}$ as

$$n'_i(z) = n_i(z - \Delta_{z,i}), \quad (30)$$

where the index $i$ runs over the tomographic redshift bins considered in the analysis. This method has been used successfully in past studies (see e.g. Troxel et al. 2018). We assume the effect of the parameters $\Delta_{z,i}$ on the cosmological constraints to be independent of cosmology and model their impact at the summary statistic level for each element $j$ of the data vector $\vec{y}$ according to

$$y_j(\theta; \Delta_{z,i}) = y_j(\theta; \Delta_{z,i} = 0)(1 + f_{i,j}(\Delta_{z,i})), \tag{31}$$

with $\theta$ denoting the remaining, fixed parameters. The functions $f_{i,j}$ are chosen as quadratic polynomials

$$f_{i,j}(\Delta_{z,i}) = c^1_{i,j}\Delta_{z,i} + c^2_{i,j}\Delta^2_{z,i}. \tag{32}$$

The coefficients of the polynomials are fitted using a set of simulations at the fiducial cosmology in which the injected redshift distributions of the source galaxies were changed according to equation (30). The simulations span the space $\Delta_{z,i} \in [-0.005, 0.005]$ in five linearly spaced points using 4000 simulations per point.

As shown in Fischbacher et al. (2022), the impact of redshift errors on cosmological constraints can be cosmology-dependent, especially when varied in combination with IA. In addition, redshift shape errors (e.g. errors of width of the redshift distributions) can also bias constraints and increase the systematic uncertainty. However, with the accuracy of the redshift estimation assumed here, these effects will play only a minor role.

### 4.2.2 Shear bias

The observed changes in the shapes of the source galaxies can be caused by other sources apart from gravitational lensing such as fluctuations in the Earth's atmosphere, instrumental effects or inaccuracies in the noise model (Hirata et al. 2004; Bernstein 2010; Melchior & Viola 2012; Refregier et al. 2012). The influence of these effects on the observed shear signal can be modelled by a multiplicative shear bias parameter $m$ and an additive shear bias parameter $\vec{c}$ as

$$\vec{\gamma}_{\rm obs} = m\,\vec{\gamma} + \vec{c}. \tag{33}$$

The impact of the additive bias on cosmological results is often found to be negligible, thanks to the extensive modelling and correction for shear biases in modern shear calibration pipelines (see e.g. Gatti et al. 2020). Hence, we do not include additive shear biases in this forecast. However, we include multiplicative shear bias parameters $m_i$ ($i$ runs over the tomographic bins in the analysis) since even small multiplicative shear biases are expected to significantly alter the outcomes of weak lensing studies (see e.g. Voigt & Bridle (2010)). Multiplicative shear bias is implemented in the simulations on the mass map level according to

$$\kappa_{m_i} = (1 + m_i)\kappa_{m_i=0}. \tag{34}$$

The effect on the data vectors is modelled in the same way as for the photometric redshift uncertainty using quadratic polynomials that are fitted based on a set of simulations covering $m_i \in [-0.005, 0.005]$ in five linearly spaced points using 4000 simulations per point.

### 4.2.3 Galaxy IA

Cosmic shear measurements rely on the assumption that the intrinsic shapes of galaxies are uncorrelated and average out to zero if the average is taken over a large enough sample of galaxies. However, this assumptions is broken in reality due to the gravitational interactions between galaxies with each others as well as the large-scale structure.

This effect, that is not accounted for in dark-matter-only simulations, is referred to as galaxy IA and can introduce strong biases in weak lensing studies (Heavens, Refregier & Heymans 2000). We model the effect of IA on the mass map level using a model developed in Fluri et al. (2019) that is based on the non-linear IA model (NLA) (Hirata & Seljak 2004; Bridle & King 2007; Joachimi et al. 2011). The method allows to obtain a pure IA signal from the PKDGRAV3 particle shells instead of a mass map by replacing the lensing kernel in equation (28) with an NLA kernel (see Z21 for details). The total mass map including the lensing and IA signal is then simply obtained by adding the two signals together $\kappa = \kappa_{\rm lens} + \kappa_{\rm IA}$. The NLA model includes three model parameters: (1) The amplitude of the overall signal $A_{\rm IA}$, (2) a parameter $\eta$ describing the redshift dependence of the signal, and (3) a parameter $\beta$, describing the dependence of the signal on the luminosity of the galaxy. We fix $\beta = 0$ and only infer $A_{\rm IA}$ and $\eta$ in the analysis, in accordance with previous studies (see e.g. Troxel et al. 2018; Zürcher et al. 2022).

Due to the strong dependency of the IA signal on the underlying cosmology we do not model the IA signal in a cosmology independent fashion like the other systematic effects. Instead we add an IA signal to each of the COSMOGRID simulations by expanding the used Sobol sequence by two dimensions ($A_{\rm IA}$, $\eta$), drawing an individual IA signal for each simulation.

### 4.3 Data compression

Depending on the number of filter scales and combined summary statistics the data vectors can contain multiple thousand entries. Although we have 16'000 simulations available at the fiducial cosmology, building the covariance matrix would result in a noisy estimate that could potentially lead to incorrect parameter constraints. Furthermore, we find that the emulator described in Section 4.4 achieves better results when trained on a compressed version $z$ of the data vectors instead of the raw data vectors $y$. Therefore, we compress the data vectors using a principal component analysis (PCA) (Pearson 1901) as in Z21 that relies on a singular value decomposition of the data. We use the `sklearn.decomposition.PCA` implementation to perform the PCA and keep enough components to explain at least 99.99 per cent of the variance of the data (Pedregosa et al. 2011). The PCA compression is built using the estimated mean data vectors $\bar{y}$ at each simulated parameter point. Hence, most of the information is contained in the first components which reduces the loss of information caused by the compression. The compressed data vectors are whitened to improve the performance of the emulator (see Section 4.4). PCA is widely used to reduce the dimensionality of long data vectors and has many applications in astronomy (see e.g. Efstathiou & Fall 1984; Yip et al. 2004; Chen et al. 2012). The dimensionality reduction happens by representing the data vectors in a lower dimensional basis of eigenvectors. These eigenvectors are found by constructing a matrix $U$ which is such that

$$U^T D U = \Lambda, \tag{35}$$

where $D$ denotes the design matrix holding the uncompressed mean data vectors $\bar{y}$ and $\Lambda$ is a diagonal matrix containing the eigenvalues of $D$. The matrix $U$ is found using singular value decomposition.

### 4.4 Emulator

Since we only have simulations of the summary statistics available for a finite set of locations in the parameter space, we train an emulator to predict the compressed data vectors for the full parameter space. Hence, we train an artificial neural network to predict the PCA components of the summary statistics from the input cosmological parameters. Since the influence of the nuisance parameters $m_i$ and $\Delta_{z,i}$ is modelled analytically, the network does not have to be trained on these parameters. The network only needs to operate on the 8D parameter space spanned by $\Omega_m$, $\sigma_8$, $n_s$, $H_0$, $\Omega_b$, $w$, $A_{IA}$, and $\eta$.

The chosen architecture consists of an input layer, three hidden, fully connected layers with 512 neurons each, as well as an output layer with a linear activation function. All hidden neurons use a Gaussian Error Linear Unit (GeLU) activation function (Hendrycks & Gimpel 2016), which we found to perform better for this problem than the more commonly used Rectified Linear Unit (ReLu) activation function. The network is trained using the Adam optimizer (Kingma & Ba 2014) with a learning rate of 0.0001 over 3000 epochs. The training data are split into mini-batches with a batch size of 32 samples each. We train an individual network for each summary statistic. We use the `tensorflow` software to implement the neural network (Abadi et al. 2015).

Before we trained the networks that are used in the actual analysis, we performed a hyperparameter optimization step to choose the learning rate, number of layers, neurons per layer, batch size, and the activation function to optimize the performance of the network.

Instead of using a standard mean-squared-error loss function we weigh the contribution of different data vector elements by the estimated variance of the element to put more emphasis on the elements that provide more constraining power

$$L_{loss} = \sum_{i=0}^{n} \frac{(z_{i,true} - z_{i,pred})^2}{\sigma_i^2}. \tag{36}$$

The predicted compressed data vector is indicated by $z_{pred}$ and the true compressed one by $z_{true}$. The sum runs over all $n$ elements in the data vector and $\sigma_i$ stands for the estimated standard deviation of the $i$-th element of the data vector at the fiducial cosmology.

To assess the performance of the emulator we select 20 randomly chosen points from the original set of 315 parameter points that are contained in the convex hull of the remaining points as the validation set. The validation loss is monitored during the training on the remaining data along with the training loss to check that the network does not overfit to the training data. We train the final networks on the whole data set due to the small number of simulations available. Hence, the obtained validation loss serves as an upper bound for the validation loss of the final network.

We judge the accuracy of the emulators as being adequate for this forecast study. However, we stress that in case of an application to real data it would be favourable to use the whole CosmoGrid simulation suite for the training. Due to the many different summary statistics and survey setups probed in this analysis this was computationally not feasible in this work.

### 4.5 Parameter inference

We infer the posterior distributions of the cosmological parameters of the $w$CDM model as well as the NLA model parameters $A_{IA}$ and $\eta$ given a compressed mock measurement of the summary statistics $\vec{z}$ using a standard Bayesian inference approach. Hence, we estimate the posterior distribution

$$p(\vec{\theta}|\vec{z}) \propto \mathcal{L}(\vec{z}|\vec{\theta})\pi(\theta), \tag{37}$$

where $\vec{\theta}$ includes the $w$CDM, NLA, multiplicative shear bias, and the redshift uncertainty parameters. The prior $\pi$ on the parameters is dictated by the size of the space that is covered by the simulations. The priors on the individual parameters are listed in Table 2. We assume the likelihood of the data to be Gaussian and we apply the same corrections as in Z21, Z22 to account for the uncertainty in the estimate of the covariance matrix as well as the simulated data vectors used to train the emulator (Sellentin & Heavens 2015; Jeffrey & Abdalla 2019)

$$\mathcal{L}(\vec{z}|\vec{\theta}) \propto \left(1 + \frac{N_\theta}{(N_\theta + 1)(N_{fid} - 1)}Q\right)^{-N_{fid}/2}, \tag{38}$$

with

$$Q = (\vec{z} - \hat{\vec{z}}(\vec{\theta}))^T \hat{\Sigma}^{-1} (\vec{z} - \hat{\vec{z}}(\vec{\theta})).$$

In the above equation, $\hat{\Sigma}$ denotes the estimate of the covariance matrix built from the $N_{fid} = 16'000$ data vector realizations at the fiducial cosmology and $\hat{\vec{z}}(\vec{\theta})$ the estimate of the data vector at a parameter location $\vec{\theta}$ estimated from $N_\theta = 560$.

We use the Markov Chain Monte Carlo (MCMC) sampler `emcee` (Daniel et al. 2013) to efficiently sample from the posterior distributions using 25 walkers per parameter (450 for the two setups with five tomographic bins and 700 for Stage 4 (10 z-bins)) with a chain length of 100'000 per walker.

## 5 RESULTS

We present a range of comparisons of the parameter constraints that we derived using the summary statistics emulators described in Section 4.4 and the parameter inference procedure outlined in Section 4.5. We start by comparing the three different survey setups and the three different filter schemes for the higher order statistics without applying any additional scale cuts. Subsequently, we study the impact of baryonic physics on the different statistics and present the necessary scale cuts for three different baryonic models. We forecast realistic constraints for a stage 4 setup using the derived scale cuts and report on the importance of galaxy IA, cross-tomographic information content, and the potential of combining different summary statistics.

It was demonstrated by Fluri et al. (2022) that the angular power spectra measured from the CosmoGrid simulations reproduce the predictions of the theory code PYCCL[16] (Chisari et al. 2019b) well for $\ell \gtrsim 20$. Hence, we do not use the lowest two $\ell$-bins in the analysis, effectively applying a lower scale cut of $\ell > 20$. For consistency, we also remove the first 4 wavelet filter scales in the DYADIC scheme in the fiducial analysis setting since they nearly exclusively contain modes with $\ell < 20$. Since the non-Gaussian information content on such large scales is expected to be very small this does not significantly alter the results. We do not modify the LOG scheme since none of its filters contain a significant portion of $\ell < 20$ modes. Alternatively, we include all modes $\ell \leq 2048$ in the fiducial angular power spectra analysis. This is approximately in concordance with the scales accessed by the higher order statistics. This is based on the observation that applying the smallest filter of the GAUSS scheme (FWHM = 3.3 arcmin) to a mass map leads to a ~50 per cent drop in power at $\ell = 2048$.

As the primary measure for the constraining power we use the Figure of Merit (FoM) that we calculate from the parameter space covariance matrix $\Sigma_{\vec{\theta}}$ according to

$$FoM(\vec{\theta}) = (|\hat{\Sigma}_{\vec{\theta}}|)^{-1/n}. \tag{39}$$

The parameter space covariance matrix is estimated from the MCMC chains and $n$ indicates the dimensionality of the parameter space.

---

[16] https://github.com/LSSTDESC/CCL

We calculate the *FoM* in the subspace that is most constrained by weak lensing measurements, namely the space spanned by $\Omega_m$, $S_8$, and $w_0$ (hence setting $n = 3$ and marginalizing over the remaining parameters). Furthermore, we also compare the widths of the 1D posterior distributions on these three parameters.

### 5.1 Comparison of filter schemes

We start by comparing how the different higher order statistics perform depending on the used filter scheme (i.e. GAUSS, LOG, and DYADIC). A visual comparison between the $w$CDM constraints obtained from the angular power spectrum analysis and the higher order statistics analyses using the most favourable filter scheme (LOG) is presented in Fig. 4. A tabular comparison in terms of FoM and 1D parameter constraints can be found in Supplementary Table C1 in Appendix C. First, we notice that the DYADIC filter scheme is outperformed by the other two schemes for all summary statistics. This is not surprising given that most of the filters in the DYADIC scheme focus on large scales on which the mass maps are close to a Gaussian random field. We also note that the DYADIC scheme was originally designed for mass map reconstruction routines and not for the extraction of non-Gaussian information from the maps and is therefore not expected to be necessarily optimal for our analysis and its relevant scales.

Comparing the GAUSS and wavelet LOG schemes we find a slight but consistent preference for the wavelet scheme. From Fig. 1, we observe that the LOG filters are more concentrated on small scales and contain less large scale modes compared to the GAUSS filters that act as low-pass filters. We suspect that this leads to better resolution of the small scale features and hence to more information extracted from the maps.

Based on these observations, we limit the discussion for each statistic to the best case scenario, namely the LOG scheme, in the following investigations.

### 5.2 Comparison of survey designs

We compare how the constraining power of the different statistics changes with the survey design. The numerical results are again included in Supplementary Table C1 in Appendix C. All statistics profit greatly from the transition from stage 3 to stage 4 (STAGE 3 → STAGE 4 (5 Z-BINS)), as expected given the increase in survey area, redshift depth, and source galaxy number density. Additionally, we find that an increase in the number of tomographic redshift bins leads to a further increase in constraining power that is observed for all statistics (STAGE 4 (5 Z-BINS) → STAGE 4 (10 Z-BINS)). We take this as an indication that the additional tomographic information can compensate for the increased shape noise in the maps caused by the lower galaxy number density per bin.

For brevity, we restrict all further investigations to the fiducial survey setup STAGE 4 (5 Z-BINS) in the following sections. We decide to use the STAGE 4 (5 Z-BINS) setup over the STAGE 4 (10 Z-BINS) setup, as we expect the results to be the same conceptually, but we are able to run more tests using the STAGE 4 (5 Z-BINS) setup due to the significantly shorter data vectors (15 compared to 55 different tomographic bin combinations). This choice is therefore motivated by computational feasibility.

### 5.3 Comparison of summary statistics

The order of the statistics with respect to constraining power is found to be peak counts, angular power spectrum, minima counts, and Minkowski functionals in descending order (see Table 3 and

Supplementary Table C1 in Appendix C). We present the constraints in the full $w$CDM parameter space in Fig. 4. We find very similar constraints for peak and minima counts with the peaks slightly outperforming the minima as it was also found in previous studies (see e.g. Coulton et al. 2020). The Minkowski functionals are outperformed by the extrema statistics in our study. While this agrees with the findings of Z21, other studies find constraints using Minkowski functionals that rival or even outperform the extrema count statistics constraints (Coulton et al. 2020).

### 5.4 Galaxy IA

We investigate the impact of galaxy IA on the different summary statistics by comparing the fiducial results to an analysis in which we fixed the galaxy IA parameters ($A_{IA} = 0$ and $\eta = 0$). The numerical results are included in Table 3. All statistics gain significantly in constraining power when galaxy IA is not considered in the analysis. We take this finding as an indication that none of the studied statistics is significantly robust to galaxy IA at the stage 4 level. The angular power spectrum constrains $A_{IA}$ the most, but also the extrema counts yield good constraints. None of the statistics can significantly constrain the redshift dependence of the galaxy IA signal $\eta$. Again, the strongest constraints are provided by the angular power spectrum. We find a slight tendency for the Minkowski functionals to be less affected by galaxy IA compared to the other statistics.

### 5.5 Tomography

We investigate how much the cross-bins (e.g. 1×2, 1×3,...) contribute to the total constraining power for the different statistics. We do so by running a set of analyses in which we do not consider the cross-bins but only the autobins. Table 3 presents a comparison of the resulting constraints compared to the fiducial results. We find that all summary statistics significantly profit from the additional cross-tomographic information. Most importantly, we find that the additional information contributes primarily to constraining galaxy IA. Without the cross-tomographic information the constraints on $A_{IA}$ worsen. This also negatively affects the cosmological constraints partially due to the strong correlations between $A_{IA}$ and the cosmological parameters (most notably $S_8$). The effect is more significant for the statistics that constrain galaxy IA stronger in the first place. Hence, we record a stronger impact for the angular power spectrum and the extrema counts.

### 5.6 Baryons

The presence of baryonic matter in cosmological simulations was demonstrated to significantly alter the distribution of matter on small scales (see e.g. van Daalen et al. 2011; McCarthy et al. 2017; Chisari et al. 2019a). Hence, baryonic effects have also been shown to affect weak lensing statistics such as the angular power spectrum (see e.g. Schneider et al. 2019), as well as peak and minima counts (see e.g. Coulton et al. 2020; Ayçoberry et al. 2022) and Minkowski functionals (Osato, Shirasaki & Yoshida 2015). Baryonic effects can be included on the level of the matter power spectrum using for example HMCODE (Mead et al. 2015). A potential way to include baryonic effects into our simulations would be to run hydrodynamical simulations. However, given that the modelling of baryonic effects, such as active galactic nuclei (AGN) feedback or stellar winds, requires the resolution of very small scales, running such simulations is computationally unfeasible for our analysis. An alternative approach was demonstrated by Tröster et al. (2019), who trained neural networks to learn to infer the distribution of gas from a given dark matter distribution. Another way to incorporate baryons into such analyses was explored by Fluri et al. (2022), who used an extension of the *baryonification model* developed in Schneider & Teyssier (2015); Schneider et al. (2019, 2021) to the shell level. Using the same method as Fluri et al. (2022) would require to increase the dimensionality of the sampled parameter space by another two parameters resulting in a very sparse sampling given the fixed number of simulations used in this study. While we hope to explore the full integration of baryons in a future study in which a larger fraction of the COSMOGRID simulations can be used, this was not computationally feasible in this work due to the large number of different survey and filter scheme configurations being probed.

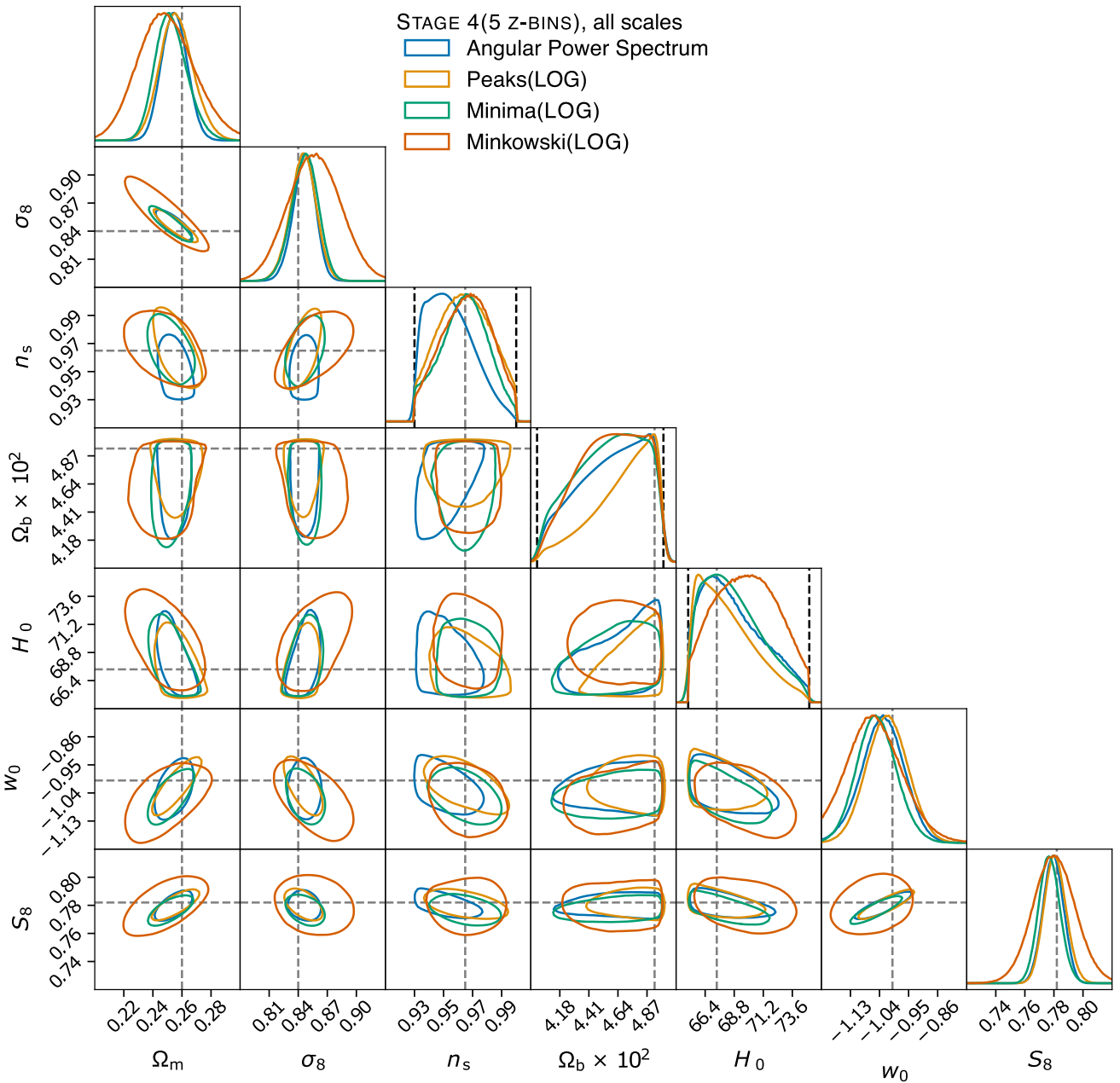

**Figure 4.** Constraints on the full $w$CDM parameter space as found in the STAGE 4 (5 Z-BINS) survey setup using the most favourable filter scheme for the higher order statistics (namely the LOG scheme). No additional scale cuts were applied. The contours show the 68 per cent confidence regions. The dashed grey lines are located at the true parameter values. We indicate the prior boundaries using a black dashed line in the panels showing the 1D constraints, if the constraints are restricted by the prior.

Instead, we explore the impact of baryons on the different summary statistics by deriving the necessary scale cuts for each individual statistic for the fiducial STAGE 4 (5 Z-BINS) setup. To do so, we follow the same approach previously used by Z22. We derive the scale cuts by studying the shift of the cosmological constraints, that is observed when swapping a mock measurement data-vector obtained from one of the simulations at the fiducial cosmology with a data-vector obtained from a baryonified version of the same base simulation with the same initial conditions. The baryonification was achieved by shifting the positions of the particles in the simulation around massive halos leading to a modification of the halo profiles to resemble realistic profiles including the gas, central galaxy and a collisionless dark matter component (see Schneider & Teyssier 2015; Schneider et al. 2019, 2021 for details). We use the same criterion for the scale cuts as Z22, requiring that the projected constraints in the $\Omega_m - S_8$ plane do not shift by more than $0.3\sigma$ when including baryons. The shift in the constraints is quantified using the `tensiometer` software developed by Raveri & Hu (2019); Raveri, Zacharegkas & Hu (2020); Raveri & Doux (2021). Contrary to Z22, who derived scale cuts using a single baryonification model, we use three baryon contaminated simulations at our fiducial cosmology that correspond to the models A-avrg, B-avrg, and C-avrg (see table 2 in Schneider et al. (2021)) and derive the scale cuts for each model. B-avrg corresponds to the best-guess model, while the models A-avrg and C-avrg potentially under- and overestimate the impact of baryons. We demonstrate how the fiducial data vector changes for the different statistics when the baryonification is applied to the simulations in Fig. 5. We find the angular power spectrum at high multipoles ($\ell > 500$) to be affected the most, with deviations reaching up to $\sim 7\sigma$ for $\ell > 1500$. For the extrema count statistics, we find the peak and minima counts to be affected similarly by the baryons with deviations reaching $\sim 5\sigma$ for high SNR peaks and low SNR minima, respectively. It was found by Coulton et al. (2020) that the minima counts are more resilient to baryons compared to the peak counts, which seems to be at odds with our findings. However, this only holds true if the high SNR tail of the peak counts is included in the comparison. Hence, this effect is not recorded in our analysis since we restrict ourselves to peaks with SNR $\leq 4$ and minima with SNR $\geq -4$. While Minkowski functionals appear to be affected less by baryons, one should take into account that their constraining power is also significantly lower, which relativizes their apparent resilience to baryonic effects.

**Table 3.** Uncertainties of the 1D-posterior distributions of the three most constrained $w$CDM parameters $\Omega_m$, $S_8$, and $w_0$, as well as the galaxy IA amplitude $A_{IA}$, and the FoM. All scales are included in the analysis. We also include the results of an analysis with fixed galaxy IA parameters ($A_{IA} = 0$, $\eta = 0$), as well as an analysis without cross-tomographic information. The uncertainties are reported as the standard deviations of the 1D posteriors. The remaining $w$CDM parameters are prior dominated and not listed here. We include the relative change of the values when IA or the cross-bins are not considered as compared to the baseline results in brackets.

| Statistic | $FoM(\Omega_m, S_8, w_0)$ | $\sigma(\Omega_m) \times 10^2$ | $\sigma(S_8) \times 10^2$ | $\sigma(w_0) \times 10^2$ | $\sigma(A_{IA}) \times 10$ |
|---|---|---|---|---|---|
| STAGE 4 (5 Z-BINS), all scales | | | | | |
| $C_\ell$ | 8413 (−) | 0.871 (−) | 0.69 (−) | 6.57 (−) | 0.72 (−) |
| Peaks (LOG) | 8936 (−) | 1.15 (−) | 0.747 (−) | 6.35 (−) | 1.6 (−) |
| Minima (LOG) | 8195 (−) | 1.1 (−) | 0.731 (−) | 6.1 (−) | 1.65 (−) |
| Minkowski (LOG) | 2347 (−) | 1.39 (−) | 2.11 (−) | 3.58 (−) | 2.76 (−) |
| STAGE 4 (5 Z-BINS) no IA, all scales | | | | | |
| $C_\ell$ | 10 943 (+ 30 per cent) | 0.673 (+ 22 per cent) | 0.542 (+ 21 per cent) | 5.38 (+ 18 per cent) | — |
| Peaks (LOG) | 13 302 (+ 48 per cent) | 0.872 (+ 24 per cent) | 0.598 (+ 19 per cent) | 5.49 (+ 13 per cent) | — |
| Minima (LOG) | 11 348 (+ 38 per cent) | 0.857 (+ 21 per cent) | 0.602 (+ 17 per cent) | 5.24 (+ 14 per cent) | — |
| Minkowski (LOG) | 3034 (+ 29 per cent) | 1.17 (+ 19 per cent) | 1.82 (+ 14 per cent) | 3.55 (+ 1 per cent) | — |
| STAGE 4 (5 Z-BINS) no cross-bins, all scales | | | | | |
| $C_\ell$ | 4959 (−41 per cent) | 1.27 (−45 per cent) | 0.899 (−30 per cent) | 8.85 (−34 per cent) | 3.82 (−430 per cent) |
| Peaks (LOG) | 4672 (−47 per cent) | 1.51 (−30 per cent) | 1.09 (−46 per cent) | 9.22 (−45 per cent) | 3.28 (−104 per cent) |
| Minima (LOG) | 4794 (−41 per cent) | 1.39 (−26 per cent) | 0.97 (−32 per cent) | 7.73 (−26 per cent) | 2.98 (−81 per cent) |
| Minkowski (LOG) | 518 (−77 per cent) | 3.22 (−132 per cent) | 3.04 (−44 per cent) | 13.4 (−275 per cent) | 9.16 (−232 per cent) |

The derived scale cuts for the different summary statistics are presented in Table 4. The imposed scale cuts are more restrictive compared to earlier studies like Z22 due to the increased constraining power of the STAGE 4 (5 Z-BINS) setup. A change in the criterion used to derive the cuts (e.g. considering the shift in $S_8$ only) could potentially change these results.

In the following Section, we apply the B-avrg scale cuts following Schneider et al. (2021) that propose the B-avrg model as the best-guess model.

### 5.7 Fiducial constraints

We present the fiducial constraints of our analysis. These constraints are tailored to represent a realistic forecast of the cosmological constraints that can be expected from a stage 4 analysis. We include the most significant weak lensing systematics (galaxy IA, multiplicative shear bias, and photometric redshift uncertainty). Realistic scale cuts are applied to prevent significant biases from unmodelled baryonic physics. The presented constraints are based on the results presented in the previous sections. Hence, we choose the following fiducial setup:

(i) filter scheme for the higher-order statistics: LOG
(ii) Survey setup: STAGE 4 (5 Z-BINS)
(iii) Scale cuts: Model B-avrg cuts

#### 5.7.1 Individual statistics

The fiducial constraints on the most constrained $w$CDM parameters ($\Omega_m, \sigma_8$ and $w_0$) are displayed in Fig. 6, jointly with the constraints on $S_8$ and $A_{IA}$. Additionally, the constraints in the full $w$CDM parameter space are presented in Supplementary Fig. B1 in Appendix B. The numerical results are included in Table 5. The strength of the constraints is strongly diminished by the scale cuts as compared to the results without cuts (see Table 3). The *FoM* is reduced by a factor of $\sim 2-3$ for the angular power spectrum, extrema counts and the Minkowski functionals. We take this significant degradation of the constraints as a strong indication that a full baryonic treatment should be included in a stage 4 weak lensing survey analysis for all summary statistics in order to avoid restrictive scale cuts and to fully realize the potential of all statistics. The ordering of the summary statistics with constraining power is modified compared to the analysis without cuts, with the angular power spectrum now slightly outperforming the peak counts. It is noteworthy that the extrema counts still provide constraints that are comparable to the angular power spectrum although most of the non-Gaussian information on small scales was removed from the data vectors by the scale cuts.

#### 5.7.2 Combined statistics

We explore different combinations of the higher order mass map statistics with the angular power spectrum. It was previously demonstrated that such combinations bear the potential to break degeneracies between parameters (Peel, Austin et al. 2018) and strongly constrain cosmology thanks to the different kinds of features and information that is extracted by the different statistics (see e.g. Liu et al. 2015; Martinet et al. 2018; Harnois-Déraps et al. 2021; Zürcher et al. 2022). Additionally, the statistics react differently to noise and systematic effects, which can improve the robustness of the constraints when combining the statistics Z21. We list the numerical results for the probed combinations in Table 5. Additionally, we present a selection of the constraints in Fig. 7. The corresponding constraints of the full $w$CDM parameter space are presented in Supplementary Fig. B2 in Appendix B.

From our findings, we conclude that already the combination of the angular power spectrum with a single higher order mass map statistic significantly increases the constraining power. Adding either peak counts or Minkowski functionals is found to be more favourable than minima counts. We note that the constraints on galaxy IA are barely improved by the addition of the higher order mass map statistics but are completely dominated by the angular power spectrum. In turn, the cosmological constraints tighten up significantly thanks to the higher order map statistics that are now more constraining on cosmology due to the galaxy IA parameters being strongly constrained by the angular power spectrum. Further, we notice that if only a single higher order statistic can be added to the angular power spectrum the addition of the Minkowski functionals is the most favourable, indicating that the topological information probed by the Minkowski functionals is more independent from the angular power spectrum information than the information probed by the extrema counts.

Adding the minima counts on top of the peak counts further improves the constraining power.

Overall, the strongest constraints are found when combining the angular power spectrum with extrema statistics plus topological statistics. While the best constraints are observed when combining all the summary statistics, it seems that the combination of the angular power spectrum with the peak counts and the Minkowski functionals is the most 'cost-effective' one, given that it only includes three statistics, but at the same time achieves a very high precision.

It is worth noticing that most of these combinations of statistics significantly outperform the angular power spectrum, even in the case when all scales up to $\ell = 2048$ are considered in the angular power spectrum only case. Hence, our results indicate that the findings of Z21, who showed that the addition of higher order statistics can compensate for the information loss caused by stringent scale cuts due to baryonic physics, are also valid at the stage 4 survey level and when considering the full $w$CDM parameter space.

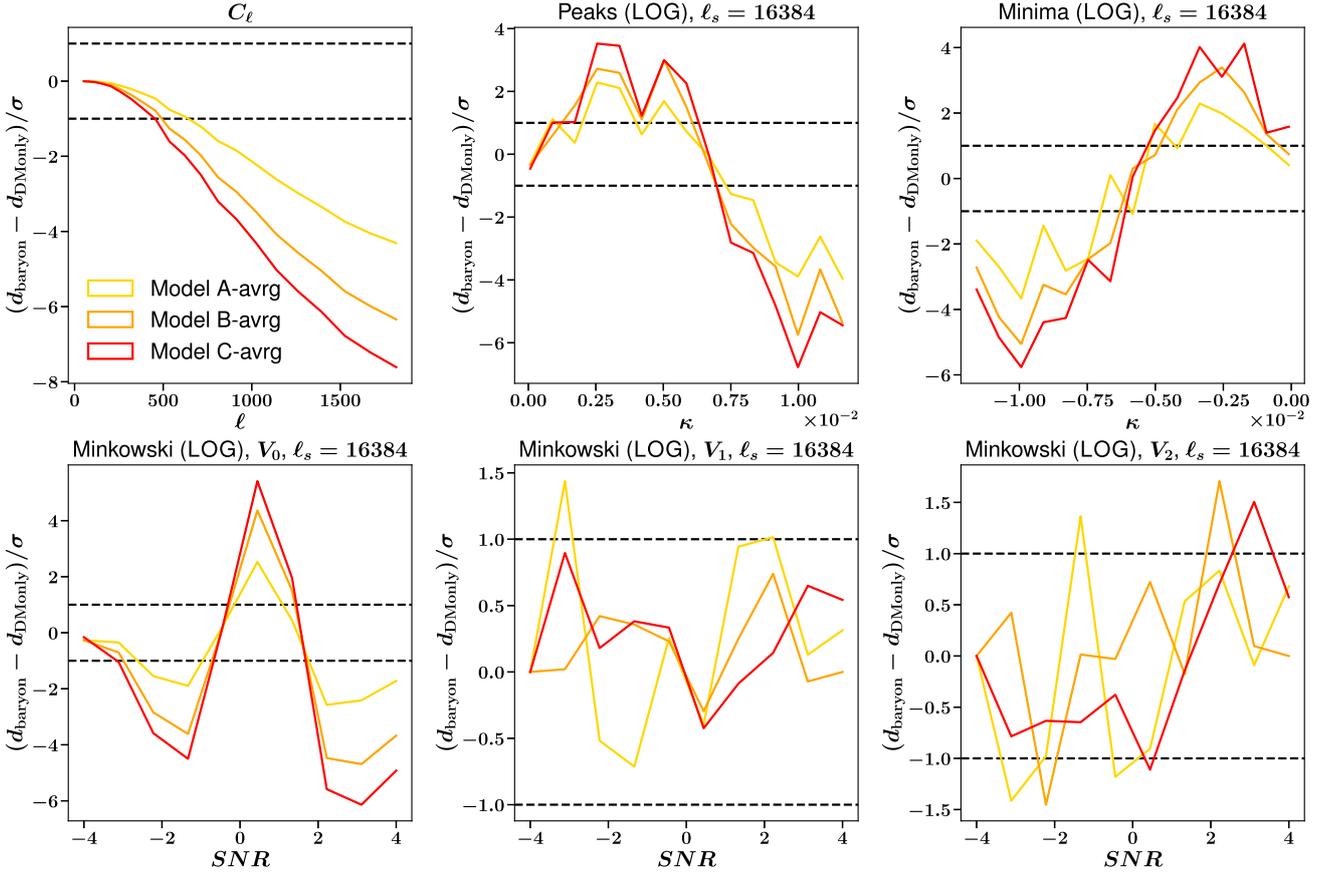

**Figure 5.** We present how the summary statistics change due to baryonic physics. Each panel shows the relative change of one of the statistics caused by baryonifying the underlying simulation using one of the three baryonic models (A-avrg, B-avrg, and C-avrg). We only present the change of the statistics for the 5×5 tomographic bin combination of the STAGE 4 (5 Z-BINS) setup. For the higher order statistics the smallest scale ($\ell_s = 16384$) is used. The relative changes are normalized by the standard deviation in each data bin that is estimated from the diagonal of the covariance matrix at the fiducial cosmology. The dashed, black lines indicate the $1\sigma$ level.

**Table 4.** We present the derived scale cuts for the different summary statistics as found for the STAGE 4 (5 Z-BINS) survey setup. The table includes the scale cuts for the three different baryonification models A-avrg, B-avrg, and C-avrg (see Table 2 in Schneider et al. (2021) for details). B-avrg corresponds to the best-guess model.

| Statistic | A-avrg | B-avrg | C-avrg |
|---|---|---|---|
| STAGE 4 (5 Z-BINS) | | | |
| $C_\ell$ | $\ell \leq 577$ | $\ell \leq 495$ | $\ell \leq 350$ |
| Peaks (LOG) | $\ell_s \leq 3315$ | $\ell_s \leq 3315$ | $\ell_s \leq 3315$ |
| Minima (LOG) | $\ell_s \leq 3315$ | $\ell_s \leq 3315$ | $\ell_s \leq 3315$ |
| Minkowski (LOG) | $\ell_s \leq 6035$ | $\ell_s \leq 3315$ | $\ell_s \leq 2223$ |

We note that, while the constraints on the remaining $w$CDM parameters ($n_s$, $\Omega_b$, and $H_0$) are prior dominated in our analysis using the individual statistics, the combined statistics approach allows to put some non-prior-dominated constraints on these parameters (see Supplementary Fig. B2 in Appendix B).

## 6 CONCLUSIONS

The ongoing stage 3 and upcoming stage 4 weak lensing surveys will allow us to map of the matter distribution in the local universe with exceptional precision and thus provide a wealth of cosmological information. To fully utilize this information for cosmological parameter inference, the use of higher order mass map statistics in addition to the more commonly used two-point statistics is essential (Petri et al. 2014). However, with the increased precision of such surveys, new theoretical as well as computational challenges arise for the analyses of such surveys.

In this work, we further improve the framework initially introduced in Z21 and successfully applied to the DES Year 3 data by Z22 to match some of the requirements posed by stage 4 surveys. Most notably, this includes the extension of the probed parameter space to the full $w$CDM space.

Furthermore, we forecast the cosmological constraints that can be expected from such surveys, using the angular power spectrum, peak counts, minima counts, and Minkowski functionals of the mass maps and we make suggestions for some analysis choices. Our main conclusions are as follows

(i) we compare different filter schemes for the higher order mass map statistics and we find a consistent preference for Starlet based filters over Gaussian filters, yielding higher constraining power. We also find, that distributing the cut-off frequencies of the Starlet filters further apart in log space instead of the originally proposed dyadic spacing is more favourable for this kind of analysis. This can be attributed to the fact, that more of the filters are located at smaller scales, while the dyadic scheme focuses more on the large scales.

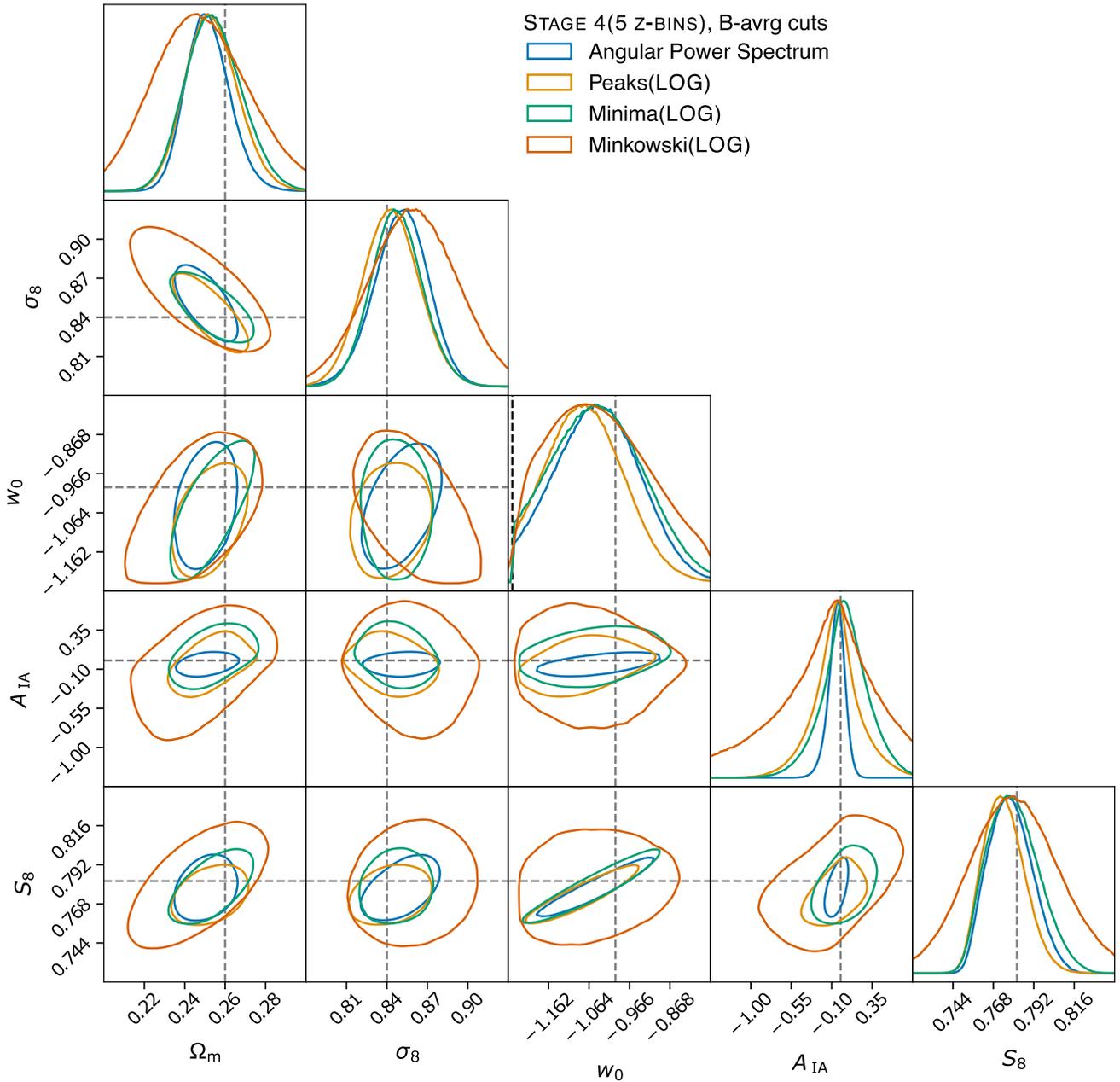

**Figure 6.** Fiducial constraints on $\Omega_m$, $\sigma_8$, $w_0$, and $S_8$, as well as $A_{IA}$ for the individual summary statistics. The presented constraints are derived for the STAGE 4 (5 Z-BINS) survey setup using the most favourable filter scheme for the higher order statistics (namely the LOG scheme). The best-guess model B-avrg scale cuts are applied (see Table 4). The presented contours indicate the 68 per cent confidence regions. The dashed grey lines are located at the true parameter values. We indicate the prior boundaries using a black dashed line in the panels showing the 1D constraints, if the constraints are restricted by the prior.

(ii) A comparison of the cosmological constraining power of the different summary statistics between a stage 4 survey setup with 5 and 10 tomographic bins revealed that all studied statistics profit from the increase in tomographic resolution.

(iii) We find the peak counts and the angular power spectrum to provide the most stringent constraints on the cosmological parameters in all tested setups. For the higher order mass map statistics peak counts achieve the highest precision, closely followed by minima counts. The extrema statistics outperform the Minkowski functionals that are geared more towards capturing the global topology of the mass maps.

(iv) We find all constraints to improve greatly if galaxy IA is not considered in the analysis. None of the statistics show significant robustness against galaxy IA for stage 4 surveys. The angular power spectrum is found to constrain the galaxy IA amplitude $A_{IA}$ the strongest.

(v) Removing the cross-tomographic bins (1×2, 1×3, etc.) from the analysis, causes the galaxy IA constraints, and hence also the cosmological constraints, to weaken considerably. This holds true for all studied summary statistics. Hence, we conclude that all statistics profit significantly from the cross-tomographic information.

**Table 5.** The uncertainties of the 1D-posterior distributions of the three most constrained $w$CDM parameters $\Omega_m$, $S_8$, and $w_0$, as well as $A_{IA}$ and the FoM for the fiducial setup. The B-avg model scale cuts were applied. The uncertainties are reported as the standard deviations of the 1D posteriors. We include the relative change of the values in comparison to the angular power spectrum results in brackets.

| Statistic | $FoM(\Omega_m, S_8, w_0)$ | $\sigma(\Omega_m) \times 10^2$ | $\sigma(S_8) \times 10^2$ | $\sigma(w_0) \times 10^2$ | $\sigma(A_{IA}) \times 10$ |
|---|---|---|---|---|---|
| STAGE 4 (5 $z$-BINS), B-avg cuts | | | | | |
| $C_\ell$ | 3921 (−) | 1.1 (−) | 1.27 (−) | 9.78 (−) | 0.941 (−) |
| Peaks (LOG) | 3362 (−14 per cent) | 1.3 (−18 per cent) | 1.21 (+4 per cent) | 8.99 (+7 per cent) | 2.74 (−191 per cent) |
| Voids (LOG) | 3039 (−22 per cent) | 1.39 (−26 per cent) | 1.49 (−17 per cent) | 10.6 (−8 per cent) | 2.68 (−185 per cent) |
| Minkowski (LOG) | 680 (−82 per cent) | 2.43 (−121 per cent) | 2.76 (−117 per cent) | 11.9 (−21 per cent) | 6.42 (−582 per cent) |
| $C_\ell$ + Peaks(LOG) | 8959 (+128 per cent) | 0.776 (+29 per cent) | 0.836 (+34 per cent) | 6.32 (+35 per cent) | 0.808 (+14 per cent) |
| $C_\ell$ + Voids(LOG) | 8171 (+108 per cent) | 0.808 (+26 per cent) | 0.936 (+26 per cent) | 6.63 (+32 per cent) | 0.894 (+4 per cent) |
| $C_\ell$ + Minkowski(LOG) | 9275 (+136 per cent) | 0.906 (+17 per cent) | 0.726 (+42 per cent) | 5.41 (+44 per cent) | 0.85 (+9 per cent) |
| $C_\ell$ + Peaks(LOG) + Voids(LOG) | 11 446 (+191 per cent) | 0.824 (+24 per cent) | 0.658 (+48 per cent) | 4.81 (+50 per cent) | 0.66 (+29 per cent) |
| $C_\ell$ + Peaks(LOG) + Minkowski(LOG) | 13 285 (+238 per cent) | 0.679 (+38 per cent) | 0.648 (+48 per cent) | 4.89 (+50 per cent) | 0.683 (+27 per cent) |
| $C_\ell$ + Voids(LOG) + Minkowski(LOG) | 11 319 (+188 per cent) | 0.76 (+30 per cent) | 0.626 (+50 per cent) | 5.08 (+48 per cent) | 0.574 (+38 per cent) |

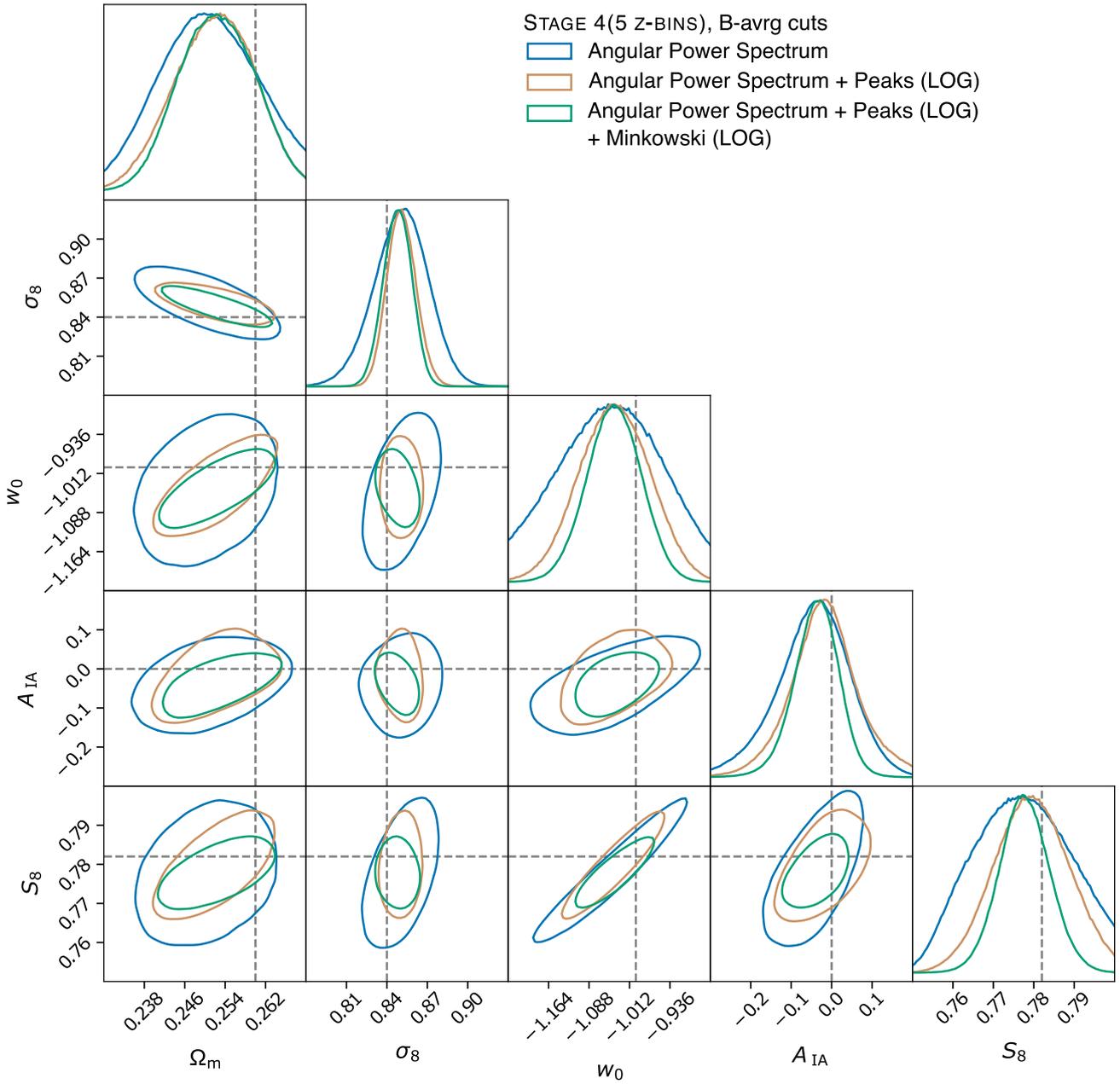

**Figure 7.** Fiducial constraints on $\Omega_m$, $\sigma_8$, $w_0$, and $S_8$, as well as $A_{IA}$ for a selection of the combined summary statistics. The presented constraints are derived for the STAGE 4 (5 Z-BINS) survey setup using the most favourable filter scheme for the higher order statistics (namely the LOG scheme). The best-guess model B-avrg scale cuts are applied (see Table 4). The contours show the 68 per cent confidence regions. The dashed grey lines are located at the true parameter values.

(vi) We derive the necessary scale cuts that need to be applied in a stage 4 scenario for the summary statistics to stay unaffected by baryonic physics, that is not modelled in dark-matter-only simulations. Due to the increased constraining power of stage 4 surveys the scale cuts are more stringent than in a stage 3 scenario leading to a significant drop of the constraining power of all summary statistics.

(vii) We find that combining multiple summary statistics has the potential to extract different, independent types of features and information from the mass maps. Combining different summary statistics is found to yield competitive constraints even when restricted by the conservative stage 4 scale cuts, enforced by unmodelled baryonic effects in the dark-matter-only simulations. Using such a combination of summary statistics provides an alternative, computationally more feasible way to obtain tight constraints on the $w$CDM model, compared to modelling small scale baryonic physics in the simulations. Additionally, we find the combined approach to mildly constrain the parameters $n_s$, $\Omega_b$, $H_0$, and $\eta$, which are prior-dominated in our analyses using the individual statistics.

While we extended the method originally proposed by Z21 to meet some of the requirements of a stage 4 survey analysis in this work, we note that there are further known and potentially also unknown systematic effects that might require additional scale cuts or modifications of the presented methodology. First, the Born approximation used in UFalcon was shown to be inaccurate for higher order statistics in a stage 4 setting (Petri et al. 2017a) and a full ray tracing algorithm should be used instead. Secondly, Eifler, Schneider & Hartlap (2009) have demonstrated that the covariance matrix varies significantly with cosmology. This effect has to be taken into account in a stage 4 scenario. A possible way to take the cosmology dependence in account was explored by Morrison & Schneider (2013) and is applicable to our methodology. We note, that the non-linear IA modelling (NLA) is approximate and does not take into account the tidal torque field. Instead, the treatment of galaxy IA should be extended to the tidal alignment and tidal torquing model (TATT) (Blazek et al. 2019). While we focused on forecasting constraints without a baryonic treatment in this study we further note that the procedure presented by Fluri et al. (2022) is applicable to our methodology and could be used to include baryonic effects into the analysis and to relax some of the scale cuts. The above listed issues will be addressed in future studies.

## ACKNOWLEDGEMENTS

The ETH Zurich Cosmology group at the Institute for Particle Physics and Astrophysics acknowledges support by grant number 200021_192243 from the Swiss National Science Foundation. We would also like to thank Uwe Schmitt from ETH Zürich for his support with the GitLab server and CI engine. Further, we would like to thank Jean-Luc Starck from CEA-Saclay for his valuable input regarding this project.

## DATA AVAILABILITY

The CosmoGrid simulations were created as a part of the Swiss National Supercomputing Centre (CSCS) Production Project 'Measuring Dark Energy with Deep Learning'. We thank Aurel Schneider for the help with baryon feedback models. Some of the results in this paper have been derived using the HEALPY and HEALPIX packages. In this study, we made use of the functionalities provided by NUMPY Walt, Colbert & Varoquaux (2011), SCIPY Virtanen et al. (2020), MATPLOTLIB Hunter (2007), tensorflow Abadi et al. (2015), and scikit-learn Pedregosa et al. (2011). We thank the main developers of CosmoStat (Jean-Luc Starck, Samuel Farrens, and François Lanusse) and Sparse2D[17] (Samuel Farrens, François Lanusse, Jean-Luc Starck, Mosè Giordano, and Antoine Grigis) for making the mentioned software packages publicly available. We thank Antony Lewis for the distribution of GetDist, on which we relied to produce some of the plots presented in this work (Lewis 2019).

[17] http://www.cosmostat.org/software/isap

# APPENDIX A: DISTRIBUTION OF THE USED COSMOGRID SIMULATIONS

We present the distribution of the 315 used CosmoGrid simulations in the sampled 8D parameter space that is spanned by the six $w$CDM parameters ($\Omega_\mathrm{m}$, $\Omega_\mathrm{b}$, $\sigma_8$, $n_\mathrm{s}$, $H_0$, and $w_0$) and the two NLA model parameters ($A_\mathrm{1A}$ and $\eta$). Each point in the parameter space is sampled by seven fully-independent PkdGrav3 simulations.

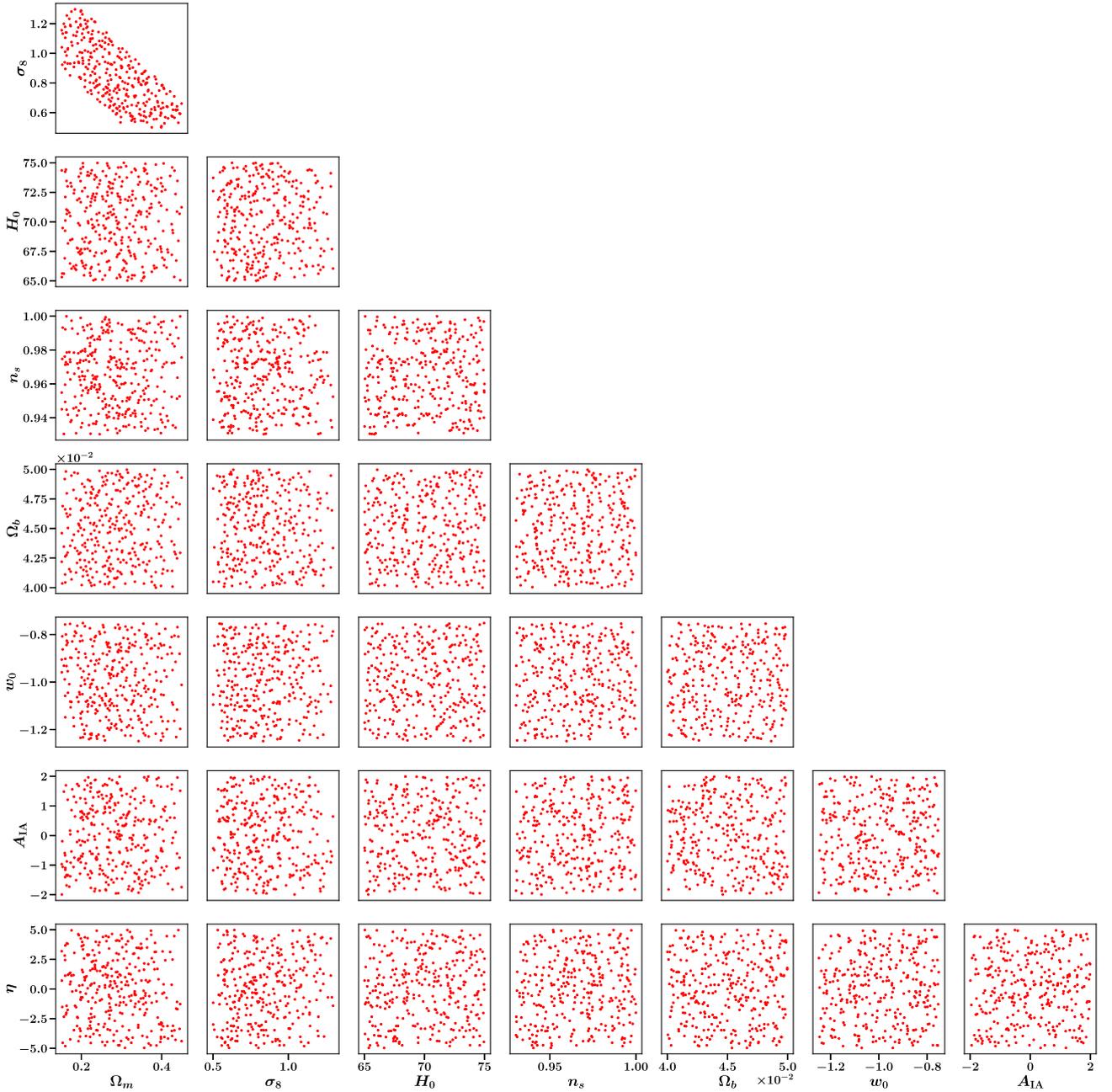

**Figure A1.** Distribution of the used CosmoGrid simulations in the full parameter space spanned by the six $w$CDM parameters ($\Omega_\mathrm{m}$, $\Omega_\mathrm{b}$, $\sigma_8$, $n_\mathrm{s}$, $H_0$, and $w_0$) and the two NLA model parameters ($A_\mathrm{1A}$ and $\eta$).

## APPENDIX B: FIDUCIAL CONSTRAINTS ON FULL *w*CDM PARAMETER SPACE

We include a visualization of the constraints on the full *w*CDM parameter space as obtained from the fiducial analysis discussed in

Section 5.7. The constraints for the individual summary statistics are shown in Fig. B1, while the contours for a selection of the combined statistics are presented in Fig. B2.

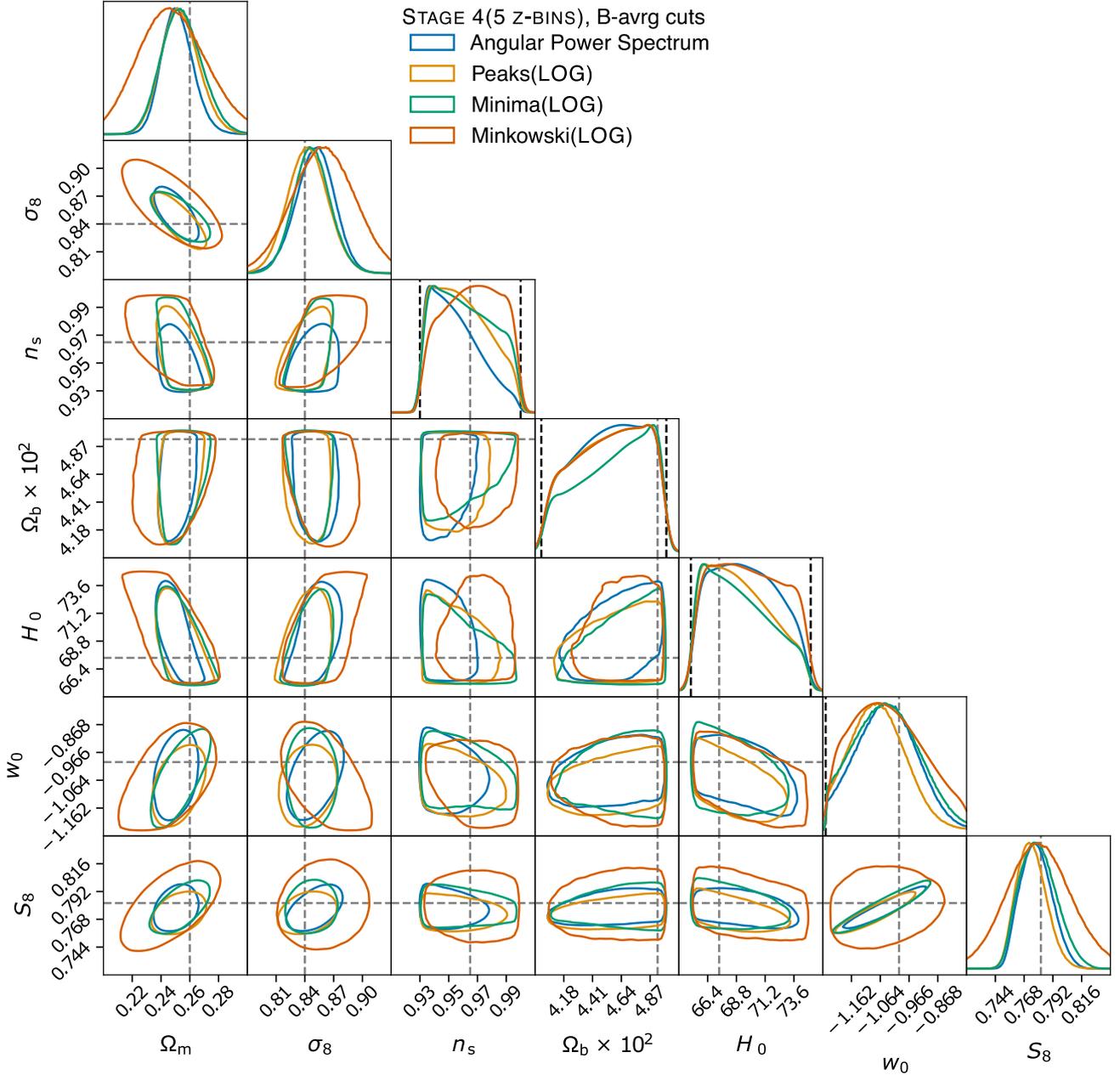

**Figure B1.** Fiducial constraints on the full *w*CDM parameter space as found in the STAGE 4 (5 Z-BINS) survey setup using the most favourable filter scheme for the higher order statistics (namely the LOG scheme). The B-avrg model scale cuts are used in this analysis. The contours show the 68 per cent confidence regions. The dashed grey lines are located at the true parameter values. We indicate the prior boundaries using a black dashed line in the panels showing the 1D constraints, if the constraints are restricted by the prior.

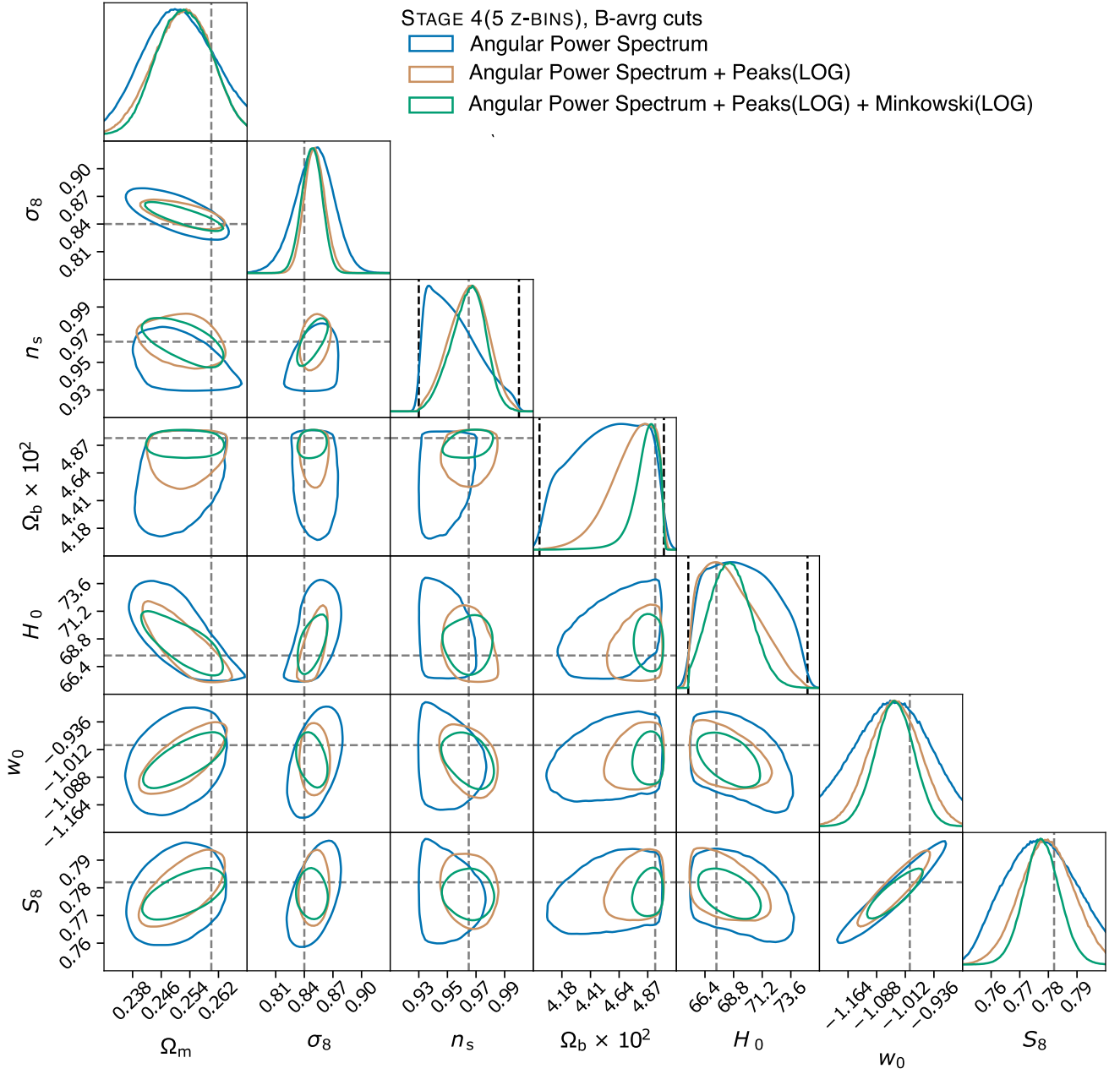

**Figure B2.** Fiducial constraints on the full *w*CDM parameter space as found in the STAGE 4 (5 Z-BINS) survey setup for a selection of the combined statistics. The B-avrg model scale cuts are used in this analysis. The contours show the 68 per cent confidence regions. The dashed grey lines are located at the true parameter values. We indicate the prior boundaries using a black dashed line in the panels showing the 1D constraints, if the constraints are restricted by the prior.

## APPENDIX C: PARAMETER CONSTRAINTS WITHOUT SCALE CUTS

We present a forecast for the constraints on the full *w*CDM parameter space. No scale cuts apart from the cuts mentioned at the beginning of Section 5 are applied. We present the derived uncertainties of the 1D-posterior distributions of the three most constrained *w*CDM parameters $\Omega_m$, $S_8$, and $w_0$, as well as the FoM of the multidimensional contours in Supplementary Table C1. The *FoM* is calculated according to equation (39) using the parameter space covariance matrix obtained from the MCMC chains. We report the uncertainties for all investigated summary statistics using the three different filter schemes (GAUSS, LOG, and DYADIC) and the three survey setups (STAGE 3, STAGE 4 (5 Z-BINS), and STAGE 4 (10 Z-BINS)).

**Table C1.** We present the uncertainties of the 1D-posterior distributions of the three most constrained $w$CDM parameters $\Omega_m$, $S_8$, and $w_0$, as well as the FoM. The uncertainties are reported as the standard deviations of the 1D posteriors. Additionally, we report on the relative change of the higher order statistics over the angular power spectrum constraints for each survey setup. No scale cuts were applied. We include the relative change of the values in comparison to the angular power spectrum results for each survey setup in brackets.

| Statistic | $FoM(\Omega_m, S_8, w_0)$ | $\sigma(\Omega_m) \times 10^2$ | $\sigma(S_8) \times 10^2$ | $\sigma(w_0) \times 10^2$ |
|---|---|---|---|---|
| STAGE 3, all scales | | | | |
| $C_\ell$ | 3471 (−) | 1.37 (−) | 0.966 (−) | 9.16 (−) |
| Peaks (GAUSS) | 3183 (−8 per cent) | 1.51 (−10 per cent) | 1.09 (−12 per cent) | 8.96 (+2 per cent) |
| Minima (GAUSS) | 2833 (−18 per cent) | 1.68 (−23 per cent) | 1.08 (−11 per cent) | 8.91 (+2 per cent) |
| Minkowski (GAUSS) | 594 (−82 per cent) | 3.4 (−148 per cent) | 2.63 (−172 per cent) | 12.1 (−32 per cent) |
| Peaks (LOG) | 4231 (+21 per cent) | 1.43 (−4 per cent) | 1.08 (−11 per cent) | 9.0 (+1 per cent) |
| Minima (LOG) | 2910 (−2 per cent) | 1.67 (−22 per cent) | 1.06 (−9 per cent) | 9.01 (+1 per cent) |
| Minkowski (LOG) | 691 (−80 per cent) | 3.43 (−150 per cent) | 2.43 (−151 per cent) | 11.4 (−24 per cent) |
| Peaks (DYADIC) | 904 (−73 per cent) | 2.0 (−45 per cent) | 2.34 (−142 per cent) | 11.7 (−27 per cent) |
| Minima (DYADIC) | 756 (−78 per cent) | 2.43 (−77 per cent) | 2.35 (−143 per cent) | 12.6 (−37 per cent) |
| Minkowski (DYADIC) | 455 (−86 per cent) | 3.3 (−141 per cent) | 3.51 (−262 per cent) | 9.41 (−2 per cent) |
| STAGE 4 (5 Z-BINS), all scales | | | | |
| $C_\ell$ | 8413 (−) | 0.871 (−) | 0.69 (−) | 6.57 (−) |
| Peaks (GAUSS) | 7502 (−10 per cent) | 1.08 (−23 per cent) | 0.777 (−12 per cent) | 6.6 (0 per cent) |
| Minima (GAUSS) | 7047 (−16 per cent) | 1.27 (−45 per cent) | 0.875 (−26 per cent) | 7.49 (−13 per cent) |
| Minkowski (GAUSS) | 1537 (−81 per cent) | 1.98 (−127 per cent) | 1.47 (−112 per cent) | 8.5 (−29 per cent) |
| Peaks (LOG) | 8936 (+6 per cent) | 1.15 (−32 per cent) | 0.747 (−8 per cent) | 6.35 (+3 per cent) |
| Minima (LOG) | 8195 (−2 per cent) | 1.1 (−26 per cent) | 0.731 (−5 per cent) | 6.1 (+7 per cent) |
| Minkowski (LOG) | 2347 (−72 per cent) | 1.39 (−59 per cent) | 2.11 (−206 per cent) | 3.58 (+45 per cent) |
| Peaks (DYADIC) | 1589 (−81 per cent) | 1.49 (−70 per cent) | 1.76 (−154 per cent) | 10.3 (−57 per cent) |
| Minima (DYADIC) | 1541 (−84 per cent) | 1.61 (−84 per cent) | 1.53 (−122 per cent) | 10.2 (−55 per cent) |
| Minkowski (DYADIC) | 1424 (−83 per cent) | 2.03 (−133 per cent) | 1.47 (−113 per cent) | 11.0 (−67 per cent) |
| STAGE 4 (10 Z-BINS), all scales | | | | |
| $C_\ell$ | 10985 (−) | 0.905 (−) | 0.681 (−) | 7.01 (−) |
| Peaks (GAUSS) | 9132 (−16 per cent) | 0.964 (−6 per cent) | 0.586 (+13 per cent) | 5.38 (+23 per cent) |
| Minima (GAUSS) | 9659 (−12 per cent) | 1.06 (−17 per cent) | 0.678 (0 per cent) | 6.2 (+11 per cent) |
| Minkowski (GAUSS) | 2102 (−80 per cent) | 1.78 (−97 per cent) | 1.3 (−90 per cent) | 6.86 (+2 per cent) |
| Peaks (LOG) | 10063 (−8 per cent) | 0.901 (0 per cent) | 0.671 (+1 per cent) | 5.61 (+19 per cent) |
| Minima (LOG) | 9969 (−9 per cent) | 1.0 (−10 per cent) | 0.633 (+7 per cent) | 6.25 (+10 per cent) |
| Minkowski (LOG) | 5462 (−50 per cent) | 1.09 (−20 per cent) | 1.34 (−97 per cent) | 1.91 (+72 per cent) |
| Peaks (DYADIC) | 1804 (−83 per cent) | 1.52 (−68 per cent) | 1.55 (−127 per cent) | 8.5 (−21 per cent) |
| Minima (DYADIC) | 2088 (−80 per cent) | 1.18 (−30 per cent) | 1.57 (−129 per cent) | 7.5 (−6 per cent) |
| Minkowski (DYADIC) | 1898 (−82 per cent) | 1.68 (−86 per cent) | 1.31 (−92 per cent) | 7.96 (−13 per cent) |

This paper has been typeset from a TeX/LaTeX file prepared by the author.